\documentclass{article}
\usepackage{amssymb}
\usepackage{graphicx} 
\usepackage{amsmath}
\usepackage{amsfonts}
\usepackage{amsthm}
\usepackage[normalem]{ulem}
\usepackage[utf8]{inputenc}
\usepackage[polish]{babel}
\usepackage{tikz}
\usetikzlibrary{decorations.pathreplacing}
\usepackage{pgfplots}
\pgfmathsetmacro{\angle}{40}
\pgfmathsetmacro{\dx}{0.5+1.2*cos(\angle)}
\pgfmathsetmacro{\dy}{1+1.2*sin(\angle)}
\pgfmathsetmacro{\shiftx}{-0.10*\dy}
\pgfmathsetmacro{\shifty}{0.10*\dx}
\usepackage{color}
\input xy
\xyoption{all} \tolerance=500
\usepackage{authblk}

\newtheorem{theorem}{Theorem}[section]

\theoremstyle{definition}
\newtheorem{example}[theorem]{Example}

\newtheorem{definition}[theorem]{Definition}

\newcommand{\R}{\mathbb{R}}

\def\sT{{\mathsf T}}

\def\xd{\mathrm{d}}
\newdir{|>}{%
!/4.5pt/@{|}*:(1,-.2)@^{>}*:(1,+.2)@_{>}}
\newdir{ (}{{}*!/-5pt/@^{(}}
\def\rel{{-\!\!\!-\!\!\rhd}}

\title{Dirac structures in nonholonomic mechanics
\footnote{The research of K.~Grabowska was partially funded by the National Science Centre (Poland) within the project WEAVE-UNISONO, No. 2023/05/Y/ST1/00043.}
}
\author{Katarzyna Grabowska}
\author{Michalina Borczyńska}
\author{Joanna Majsak}
\author{Tomasz Sobczak}
\affil{Faculty of Physics, University of Warsaw}

\pgfplotsset{compat=1.18} 

\begin{document}

\maketitle

\begin{abstract}
The concept of a Dirac algebroid, which is a linear almost Dirac structure on a vector bundle, was designed to generate phase equations for mechanical systems with linear nonholonomic constraints. We apply it to systems with magnetic-like or gyroscopic potentials, that were previously described by means of almost Poisson structures. The almost Poisson structures present in the literature in this context were constructed using constraints, metrics and information about magnetic or gyroscopic potential present in the Hamiltonian function of the system. The Dirac algebroid we use is constructed out of constraints and canonical geometric structures of the underlying bundles and is universal in the sense that it is independent on the particular Hamiltonian or Lagrangian. We provide examples showing that using the same Dirac structure we can describe systems with different potentials, magnetic or mechanical, added freely to a function generating the dynamics. 

\end{abstract}

\section{Introduction}

In the literature, the term {\it Nonholonomic Mechanics} usually refers to the study of classical mechanical systems subjected to constraints posed on velocities and described by a Lagrangian function defined on positions and velocities. From the variational point of view, classifying constraints, i.e. into {\it holonomic}, {\it nonholonomic}, {\it vakonomic}, or others, requires more than just information about properties of the admissible infinitesimal configurations; we also need to know the set of admissible variations. Here, we discuss a geometric approach rather than variational, therefore we shall adopt the point of view according to which a nonholonomic mechanical system with linear constraints is given by a manifold $M$ of positions, a possibly non-involutive distribution $C$ on $M$ representing admissible velocities, and a Lagrangian function defined on the tangent bundle $\sT M$, namely
$$L:\sT M\longrightarrow \R.$$
The dynamics of the system is determined by the constrained Euler-Lagrange equations of the form
$$\left\{\begin{array}{l}\vspace{5pt}
\displaystyle\frac{d}{dt}\left(\frac{\partial L}{\partial \dot q^i}\right)-\frac{\partial L}{\partial q^i}=\lambda_A\frac{\partial \Phi^{A}}{\partial \dot q^i},  \\
\Phi^A(q^i, \dot q^j)=0.
\end{array}\right.$$
Here, $(q^i)$ are coordinates on the manifold $M$, $(q^i,\dot q^j)$ are the adapted coordinates on the tangent bundle $\sT M$, the constraints $C$ are given by equations $\Phi^A(q^i,\dot q^j)=0$, and $\lambda_A$ denote Lagrange multipliers.

We aim to provide a universal geometric approach to such a system. In particular, we would like to be able to define a phase space and provide phase equations. We are motivated by the recent article \cite{Garcia2023} in which the case of a system with nonholonomic constraints and a Lagrangian with magnetic-like terms is discussed.
\medskip

In non-constrained cases, we can describe a mechanical system not only in a variational way, deriving Euler-Lagrange equations. Instead, we can use one of the procedures that are collectively known as Geometric Mechanics. We can use the Hamiltonian approach, according to which the dynamics is given by a system of first-order differential equations on curves in a symplectic phase space. The phase space, i.e., the space of positions and momenta, is usually the cotangent bundle $\sT^\ast M$ together with the canonical symplectic form $\omega_M$. Phase equations are encoded in the Hamiltonian vector field $X_H$ associated with a Hamiltonian function $H:\sT^\ast M\rightarrow \R$ via the formula
$$\imath_{X_H}\omega_M=-dH.$$
In this approach, Euler-Lagrange equations are differential consequences of the phase equations. More generally, Hamiltonian mechanics can be formulated employing a Poisson structure expressed as a Poisson bracket of functions or, equivalently, as a Poisson bivector.

In some cases, the dynamics can also be given by a certain vector field on the space $\sT M$ of infinitesimal configurations. In this approach, we define the Legendre map $\lambda:\sT M\rightarrow \sT^\ast M$ by means of a vertical differential of the Lagrangian. In coordinates this map reads $\lambda(q^i,\dot q^j)=(q^i, \frac{\partial L}{\partial \dot q^j})$. We call a Lagrangian function {\it hyperregular} when the Legendre map is a diffeomorphism. In such a case, we can define the symplectic form $\omega_L=\lambda^\ast(\omega_M)$ and use the energy function $E(v)=\langle\lambda(v),v\rangle-L(v)$ as a function generating the Hamiltonian vector field $X_E$,
$$\imath_{X_E}\omega_L=-dE.$$
It is easy to check that, again, Euler-Lagrange equations are differential consequences of the equations encoded in the vector field $X_E$.
\smallskip

In the following, we will use the Tulczyjew approach to Geometric Mechanics, known as the {\it Tulczyjew triple}, and introduced in the series of papers \cite{WMT76a, WMT76b, WMT77}. It is described in detail in Section \ref{sec:2}. It is the most general approach that does not put any restrictions on Lagrangians. It serves both for Hamiltonian and Lagrangian formulations of dynamics and, moreover, it is relatively easy to generalize to the constrained case. Such a generalization was proposed in \cite{GRABOWSKA2011}. The structure used to produce the generalized triple was called a {\it Dirac algebroid}, which is a linear almost Dirac structure on a vector bundle. Using Dirac algebroids means that we have to work with relations instead of maps and often apply integrability conditions on the dynamics.   

Note, that the concept of a Dirac structure was first proposed by Dorfman in \cite{Dorfman1987} in the context of integrable evolution equations and then defined geometrically by Courant in \cite{Courant1990}. The idea of using the structure in Lagrangian mechanics is due to Yoshimura and Marsden \cite{YoshimuraMarsden2006a,YoshimuraMarsden2006b}, who used Dirac and almost Dirac structures to work with Lagrangian systems with Lagrangians that are not regular. Our approach, developed in \cite{GRABOWSKA2011}, is based on linear Dirac structures, double vector bundles and the Tulczyjew approach to Geometric Mechanics. 

In the literature, several other procedures to treat nonholonomic constraints were proposed. They usually have limitations in the sense that they require additional structure to be present, e.g., the configuration space has to be a Riemannian manifold, or there are restrictions on the type of Lagrangians that can be used. As examples of such procedures, we present two constructions allowing for both Lagrangian and Hamiltonian formulation of nonholonomic mechanics for special classes of Lagrangians and Hamiltonians.

In example \ref{ex:1}, following \cite{Grabowski2008}, we discuss the case of a mechanical Lagrangian defined on the tangent bundle of a Riemannian manifold $(M,g)$. A Lagrangian is then given as the difference between the kinetic energy, quadratic in velocities, and some potential $V$ depending only on positions. The presence of a metric, together with the special form of a Lagrangian, makes it possible to define a skew algebroid structure on the linear constraints $C$ being a subbundle of $\sT M$. The skew algebroid structure is then used to construct first-order differential equations on the phase space out of the Lagrangian or Hamiltonian. The phase space in this case is $C^\ast$ -- the dual of the constraint subbundle. To illustrate this construction, we derive phase equations for a skate on a tilted ice-rink in a gravitational field.

Next, in example \ref{ex:2}, we describe the construction of an almost Poisson structure on $C^\ast$, defined in \cite{Garcia2023}, appropriate for the case when a Lagrangian contains a magnetic-like term linear in velocities. The structure is affine, therefore it does not correspond to an algebroid on $C$. As an illustration for this construction, we also use a skate on an ice-rink, but this time we put a charge on the skate and switch on a magnetic field.
\medskip

In both cases, the dynamics is indeed Hamiltonian with respect to a certain skew-symmetric bracket on $C^\ast$. In the first case, the bracket is defined by the constraint $C$ and the metric $g$, and in the second - by the constraint, metric, and Lagrangian, because the magnetic term enters partially into the bracket. Both constructions are correct in the sense that they give proper dynamical equations, however, they can be applied only for restricted classes of Lagrangians. Moreover, in the case of a Lagrangian with a magnetic term, the geometric structure depends on the Lagrangian itself, which we want to avoid.
\medskip

In section \ref{sec:3}, we recall the approach to nonholonomic mechanics proposed in \cite{GRABOWSKA2011} and based on a particular example of a Dirac algebroid associated with constraints. This approach can be used for any Lagrangian and any Hamiltonian function, however, the tools used are very different from the customary brackets, forms, or bivectors. We show that the Dirac approach gives the same dynamics in the cases discussed in examples \ref{ex:1} and \ref{ex:2}, but, in principle, is more general. In particular, the same Dirac structure defined by $C$ serves for both examples. The case of a skate, both with and without charge, is discussed within the Dirac structure approach to facilitate the comparison of different structures. In a final example we discuss the case of a charged ball rotating without slipping, in a magnetic field, with a harmonic potential.

\section{Tools of Geometric Mechanics}\label{sec:2}

Let us start with setting up basic notions and notation. Our manifolds will be paracompact and smooth. For a tangent and cotangent bundle we will use, respectively, symbols $\tau_M:\sT M\rightarrow M$ and $\pi_M:\sT^\ast M\rightarrow M$. Coordinates $(q^i)$, defined on an open subset $U$ of a manifold $M$, allow us to define adapted coordinates $(q^i, \dot q^j)$ in $\tau_M^{-1}(U)$ by using $(\partial_{q^i})$ as a basis of sections of the tangent bundle. An element $v\in \sT M$ can be therefore written in coordinates as $\dot q^i(v)\partial_{q^i}$. Similarly, $(q^i, p_j)$ are coordinates in $\pi^{-1}_M(U)$ corresponding to the basis $(\xd x^i)$; a covector $\phi$ in coordinates reads $p_i(\phi)\xd q^i$.

Tangent and cotangent bundles both come with canonical structures. The total space of the cotangent bundle $\sT^\ast M$ is a symplectic manifold with respect to the symplectic form $\omega_M=\xd \theta_M$. The one-form $\theta_M$, called {\it tautological} or {\it Liouville}, is given by the formula
$$\theta_M(w)=\langle\tau_{\sT^\ast M}(w),\, \sT\pi_M(w)\rangle,$$
for $w\in \sT\sT^\ast M$. One can see that it is indeed a canonical object. In coordinates, $\theta_M$ reads $\theta_M=p_i\xd q^i$, which explains the name tautological. Consequently, $\omega_M=\xd p_i\wedge \xd x^i$. The canonical symplectic form is linear, i.e., it is compatible with the vector bundle structure of the cotangent bundle. Technically, it means that $h_t^\ast \omega_M=t\omega_M$, where $h_t(\phi)=t\phi$ is the usual operation of multiplying a covector by a number.

The symplectic structure can be used to produce Hamiltonian vector fields for smooth functions defined on $\sT^\ast M$. For a function $H:\sT^\ast M\rightarrow \R$, the Hamiltonian vector field $X_H$ is defined by $\imath_{X_H}\omega_M=-\xd H$, which in coordinates gives
$$X_H=\frac{\partial H}{\partial p_i}\partial_{q^i}-\frac{\partial H}{\partial q^j}\partial_{p_j}.$$
Instead of using the two-form $\omega_M$, we can use the corresponding Poisson bivector $\Lambda_M$ such that $$\omega_M(X_f,X_g)=\Lambda_M(\xd f,\xd g).$$
In coordinates $(q^i, p_j)$, we have
$$\Lambda_M=\partial_{p_i}\wedge\partial_{q^i}.$$
In the context of the Tulczyjew triple, we usually use the isomorphism
$$\beta_M:\sT\sT^\ast M\longrightarrow \sT^\ast\sT^\ast M,$$
defined by $\omega_M$, namely
\begin{equation}\label{eq:0}\beta_M(w)=\omega_M(\cdot, w).\end{equation}
If in $\tau^{-1}_{\sT^\ast M}(\pi^{-1}_M(U))$ we use adapted coordinates $(q^i, p_j, \dot q^k, \dot p_l)$ and, similarly, in $\pi^{-1}_{\sT^\ast M}(\pi^{-1}_M(U))$ coordinates $(q^i, p_j, u_k, w^l)$, we can write $\beta_M$ in the form
\begin{equation}\label{eq:18}(q^i, p_j,u_k,w^l)\circ\beta_M=(q^i,p_j,-\dot p_k, \dot q^l).
\end{equation}
Note that coordinates $u_k$ and $w^l$ are defined in such a way that an element $\psi\in\sT^\ast\sT^\ast M$ reads $\psi=u_k(\psi)\xd q^k+ w^{l}(\psi)\xd p_l$.  Both $\Lambda_M$ and $\beta_M$ are compatible with the vector bundle structure of the cotangent and tangent bundles. For $\Lambda_M$ it means that
$$\mathcal{L}_{\nabla_{\sT^\ast M}}\Lambda_M=-\Lambda_M,$$
where $\nabla_{\sT^\ast M}$ is the Liouville vector field on $\sT^\ast M$, in cordinates $\nabla_{\sT^\ast M}=p_i\partial_{p_i}$. The compatibility of $\beta_M$ means that it is a double vector bundle morphism of $\sT\sT^\ast M$ into $\sT^\ast\sT^\ast M$. Moreover, it is a diffeomorphism and an anti-symplectomorphism between $(\sT\sT^\ast M, \xd_{\sT}\omega_M)$ and $(\sT^\ast\sT^\ast M, \omega_{\sT^\ast M})$. By $\xd_{\sT}\omega_M$ we denote the tangent lift of the canonical symplectic form $\omega_M$. In adapted coordinates, $\xd_{\sT}\omega_M=\xd \dot p_i\wedge\xd q^i+\xd p_j\wedge \xd \dot q^j$.
\medskip

The canonical structure of the tangent bundle $\sT M$ is that of a {\it Lie algebroid}.

\begin{definition}\label{d:1}
A Lie algebroid is a vector bundle $\tau:E\rightarrow M$ together with a Lie bracket $[\,\cdot\,,\,\cdot\,]: \mathsf{Sec}(\tau)\times\mathsf{Sec}(\tau)\rightarrow \mathsf{Sec}(\tau)$ and a vector bundle morphism $\rho:E\rightarrow \sT M$ covering identity on $M$, called the anchor, and such that
$$[X,fY]=f[X,Y]+(\rho(X)f)Y$$
for any $X,Y\in \mathsf{Sec}(\tau)$ and $f\in\mathcal{C}^\infty(M)$.
\end{definition}

Canonical examples of Lie algebroids are the tangent bundle $\sT M$, together with the usual Lie bracket of vector fields and the identity on $\sT M$ as an anchor, and a Lie algebra treated as a vector bundle over one point manifold, together with the Lie algebra bracket and the zero map as an anchor.

As a matter of fact, the Poisson bivector $\Lambda_M$ and the Lie bracket of vector fields, i.e., the Lie algebroid structure of $\sT M$ are equivalent in the sense that one defines the other. Any vector field $X$ on $M$ defines a fiberwise linear function $\imath_X$ on $\sT^\ast M$ by the formula
$$\imath_X(\phi)=\langle\phi, X(\pi_M(\phi))\rangle.$$
The canonical Poisson structure on $\sT^\ast M$ is linear, therefore, it is determined by the bracket of linear functions and functions constant on fibers. The correspondence between the bracket of vector fields and the Poisson structure is given by
\begin{equation}\label{eq:15}
\{\imath_X,\imath_Y\}=\imath_{[X,Y]}, \quad \{\imath_X, \pi_M^\ast f\}=\pi_M^\ast (X f),
\end{equation}
where $f$ is a smooth function on $M$.

The equivalence between a Lie algebroid and linear Poisson structures does not occur only on a pair of bundles $\sT M$ and $\sT^\ast M$. If a vector bundle $\tau: E\rightarrow M$ is equipped with a Lie algebroid structure, then the dual $\pi: E^\ast \rightarrow M$ carries a corresponding linear Poisson bracket. The formulae are as in (\ref{eq:15}), where we replace vector fields by sections of $\tau$. The Poisson bracket on functions on $E^\ast$ obtained this way can be represented by a linear bivector $\Lambda$, or a double vector bundle morphism
$$\Lambda^\#: \sT^\ast E^\ast\longrightarrow \sT E^\ast,\quad \Lambda^\#(\varphi)=\imath_\varphi \Lambda,$$
replacing $\beta_M$ in this more general situation.
\smallskip

Another important element of the geometry is the canonical isomorphism $\mathcal{R}_E$ of double vector bundles $\sT^\ast E^\ast$ and $\sT^\ast E$. The graph of this isomorphism is the Lagrangian submanifold generated in $\sT^\ast(E\times E^\ast)$ by the function
$$E\times_M E^\ast\ni (v,\phi)\longmapsto \langle\phi, v\rangle\in\R$$
defined on the submanifold $E\times_M E^\ast$ of $E\times E^\ast$. To identify the Lagrangian submanifold with the graph of a map $\mathcal{R}_E$, one should treat $\sT^\ast(E\times E^\ast)$ with the canonical symplectic form of a cotangent bundle as $\sT^\ast E\times \sT^\ast E^\ast$ with $\omega_{E}\ominus\omega_{E^\ast}$.

Composing $\mathcal{R}_E$ with $\Lambda^\#$, we get the map
\begin{equation}\label{eq:14}
\varepsilon:\sT^\ast E\longrightarrow \sT E^\ast, \quad \varepsilon=\Lambda^\#\circ \mathcal{R}_E,
\end{equation}
which is another double vector bundle morphism that carries the complete information about the Lie algebroid structure we started with.
\medskip

Summarising the above presentation, we can state that a Lie algebroid structure on a vector bundle $\tau: E\rightarrow M$ can be equivalently represented as a linear Poisson structure on $E^\ast$, or one of the double vector bundle morphisms
$$\varepsilon: \sT^\ast E\longrightarrow \sT E^\ast$$
or
$$\Lambda^\#: \sT^\ast E^\ast\longrightarrow \sT E^\ast,$$
covering identity on $E^\ast$ (and fulfilling a few more conditions equivalent to $\Lambda$ being a Poisson bivector).  In the case of $E=\sT M$, we have the canonical examples of these structures: the canonical Lie bracket of vector fields as a Lie algebroid, the canonical symplectic or Poisson structure on $\sT^\ast M$, as well as the double vector bundle isomorphism $\beta_M$. Note that, in this case, $\beta_M=(\Lambda_M^\#)^{-1}$. There is also the canonical version of $\varepsilon$, the {\it Tulczyjew isomorphism} $\alpha_M$, that we describe in more detail in the next section, together with its application in Geometric Mechanics.

\subsection{The classical Tulczyjew triple}\label{sec:2_1}

As we have already declared, throughout this work we will use the Tulczyjew approach to Geometric Mechanics, the symbol of which is the classical Tulczyjew triple. It is a pair of double vector bundle isomorphisms $\alpha_M$ and $\beta_M$ constituting the following commutative diagram
$$\xymatrix{
\sT^\ast\sT^\ast M\ar[dr]^{\pi_{\sT^\ast M}} & &  \sT\sT^\ast M\ar[ll]_{\beta_M}\ar[rr]^{\alpha_M}\ar[dl]_{\tau_{\sT^\ast M}}\ar[dr]^{\sT\pi_M} & & \sT^\ast\sT M\ar[dl]_{\pi_{\sT M}} \\
 & \sT^\ast M\ar[dr]^{\pi_M} & & \sT M\ar[dl]_{\tau_M} &  \\
& & M & &
}
$$

The right-hand side of the triple is a base for a Lagrangian description of an autonomous mechanical system with the configuration manifold $M$ and the phase manifold $\sT^\ast M$. The double vector bundle isomorphism $\alpha_M$ is usually defined as a dual morphism to the canonical flip $\kappa_M:\sT\sT M\rightarrow \sT\sT M$. It turns out to be the inverse map to the morphism representation of the canonical Lie algebroid structure on $\sT M$ as in (\ref{eq:14}). It is interesting that the map $\alpha_M$ also has a variational interpretation. One can find the details in \cite{Tulczyjew}.

The phase equations, called in this context {\it the dynamics}, form a Lagrangian submanifold of the symplectic manifold $(\sT\sT^\ast M, \xd_{\sT}\omega_M)$.  The dynamics is given as $\mathcal{D}_L=\alpha_M^{-1}(\xd L(\sT M))$. Depending on the properties of a particular Lagrangian, the dynamics may be in explicit form, i.e., equal to the image of a vector field on the phase space, or it may also be an implicit differential equation. The structure of $\alpha_M$ as a double vector bundle morphism is illustrated by the following commutative diagram. The dotted lines represent identity maps of the respective manifolds. The only not obvious map in the diagram below is $\xi_{\sT M}:\sT^\ast\sT M\rightarrow \sT^\ast M$. It amounts to restricting a covector on $\sT M$ to the subspace of vertical vectors tangent to $\sT M$. This subspace can be identified with an appropriate fiber of $\sT M$ itself. 
$$\xymatrix@C-10pt{
{\color{red} \mathcal{D}}\ar@{ (->}[r] & \sT\sT^\ast M\ar[dr]^{\sT\pi_M}\ar[ddl]^(.65){\tau_{\sT^\ast M}}\ar[rrr]^{\alpha_M} & & &
\sT^\ast\sT M\ar[dr]_{\pi_{\sT M}}\ar[ddl]^(.65){\xi_{\sT M}} & \\
 & & \sT M\ar@{.}[rrr]\ar[ddl]_(.35){\tau_M} & & & \sT M\ar[ddl]_(.35){\tau_M}\ar@/_1pc/[ul]_{\color{red}d L}  \\
 \sT^\ast M\ar@{.}[rrr]\ar[dr]^{\pi_M} & & & \sT^\ast M\ar[dr]^{\pi_M}^{\pi_M} & & \\
 & M\ar@{.}[rrr] & & & M &
 }$$

We have already introduced local coordinates $(q^i, \dot q^j)$ in the tangent bundle and $(q^i, p_j)$ in the cotangent bundle. In the iterated tangent and cotangent bundles, we can also use adapted coordinates. Following the rule that we add dots over coordinate symbols when applying the tangent functor, we get $(q^i, p_j,\dot q^k, \dot p_l)$ -- coordinates that we have used for $\beta_M$ in the previous section. On $\sT^\ast\sT M$, we shall use the coordinates
\begin{equation}\label{eq:16}
(q^i, \dot q^j, a_k, b_l),
\end{equation}
where first two sets of coordinates are pulled back from $\sT M$, and $a_k$ and $b_l$ correspond to the basis composed of one-forms $\xd q^k$ and $\xd \dot q^l$, respectively. The Tulczyjew isomorphism $\alpha_M$ reads in these coordinates
\begin{equation}\label{eq:16a}
(q^i,\dot q^j, a_k, b_l)\circ\alpha_M=(q^i, \dot q^j, \dot p_k, p_l).
\end{equation}

Since we are going to discuss linear constraints posed on velocities, it will be convenient to use other coordinates in the tangent and cotangent bundles, adapted to the constraints. Let $C$ denote a subbundle of the tangent bundle of the manifold $M$, for simplicity, we shall assume that it is supported on the whole base manifold. Instead of the local bases $(\partial_{q^i})$ and $(\xd q^i)$, we shall use a pair of the dual bases, $(f_i)$ -- sections of $\sT M $ and $(\phi^j)$ -- sections of $\sT^\ast M$, such that $C_q=\mathsf{Span}\{f_1,\ldots f_k\}$ and $C^\circ_q=\mathsf{Span}\{\phi^{k+1},\ldots \phi^n\}$. 
Coordinates associated with $(f_i)$ will be denoted as $(x^i)$, and sometimes split into groups $(x^a, x^\alpha)$ with $a\in\{1,\ldots, k\}$, $\alpha\in\{k+1,\ldots, n\}$. Similarly, coordinates associated with $(\phi^i)$ will be denoted as $(\eta_i)$, and split into groups $(\eta_a, \eta_\alpha)$. In the iterated tangent and cotangent bundles, we shall use the usual constructions to obtain coordinates
\begin{equation}\label{eq:3}(q^i, \eta_a,\eta_\alpha, \dot q^k, \dot \eta_b, \dot\eta_\beta) \text{ in } \sT\sT^\ast M\end{equation}
and
\begin{equation}\label{eq:17}(q^i, x^a, x^\alpha, a_j, \xi_{b}, \xi_{\beta})\text{ in }\sT^\ast \sT M.
\end{equation}
Instead of writing $\alpha_M$ in these new coordinates, we shall rather write $\varepsilon_M=\alpha_M^{-1}$, because this is the map we shall use later on:
\begin{equation}\label{eq:12}
(\,q^i, \eta_a,\eta_\alpha, \dot q^k, \dot \eta_b, \dot\eta_\beta\,)\circ \varepsilon_M=
(\,q^i, \xi_a,\xi_\alpha, \rho^k_jx^j,\, c^{l}_{bj}\eta_{l}x^j+\rho^{l}_ba_{l},\, c^{l}_{\beta j}\xi_{l}x^j+\rho^{l}_\beta a_{l}\,).
\end{equation}
In the above formula, indices $i,j,k,l$ go from $1$ to $m$, indices $a,b$ from $1$ to $k=\dim C$, and $\alpha,\beta$ from $k+1$ to $m$. Moreover, $\rho$ denotes the matrix of coordinate change, i.e. $\dot q^i=\rho^i_ax^a+\rho^i_\alpha x^\alpha$, and functions $c^i_{jk}$ describe the Lie bracket of basic sections $[f_k, f_j]=c^i_{jk}f_i$. If the functions $c^\alpha_{bd}$ are not equal to zero, then the subbundle $C$ is not integrable, therefore it is not a Lie subalgebroid of the tangent bundle, which is usually the case for nonholonomic constraints. The functions $c^i_{jk}$ can be calculated from the components of the matrix $\rho$ according to the formula
$$c^i_{jk}=(\partial_l\rho^n_j)\rho^l_k(\rho^{-1})^i_n-(\partial_l\rho^n_k)\rho^l_j(\rho^{-1})^i_n.$$
\medskip

The Hamiltonian side of the Tulczyjew triple is based on the map $\beta_M$ defined by means of the symplectic form $\omega_M$ (see (\ref{eq:0})), so the dynamics associated to a Hamiltonian function $H$ can be written as $\beta_M(\xd H(\sT^\ast M))$. This time the dynamics is always explicit, i.e., it is the image of the Hamiltonian vector field $X_H$ associated with $H$.
The appropriate diagram reads
$$\xymatrix@C-10pt{
 & \sT^\ast\sT^\ast M\ar[dr]^{\xi_{\sT^\ast M}}\ar[ddl]^(0.65){\pi_{\sT^\ast M}} & & & \sT\sT^\ast M\ar[lll]_{\beta_M}\ar[dr]^{\sT\pi_M}\ar[ddl]^(.65){\tau_{\sT^\ast M}} & \\
 & & \sT M\ar@{.}[rrr]\ar[ddl]_(.35){\tau_M} & & & \sT M\ar[ddl]_(.35){\tau_M} \\
 \sT^\ast M\ar@{.}[rrr]\ar[dr]^{\pi_M}\ar@/^1pc/[uur]^{\color{red}d H} & & & \sT^\ast M\ar[dr]^{\pi_M}^{\pi_M}\ar@/^1pc/[uur]^{\color{red}X_H} & & \\
 & M\ar@{.}[rrr] & & & M &
}.$$
Again, the only not obvious map in the diagram above is $\xi_{\sT^\ast M}:\sT^\ast\sT^\ast M\rightarrow \sT M$. It is defined as $\xi_{\sT M}$ by restricting a covector on $\sT^\ast M$ to the subspace of vertical vectors tangent to $\sT^\ast M$. 
The expression for $\beta_M$ in coordinates adopted from $M$ was already given in formula (\ref{eq:18}).
Coordinates $(q^i, \eta_a,\eta_\alpha)$ on $\sT^\ast M$, adapted to $C$, give rise to the coordinates
\begin{equation}\label{eq:4}(q^i, \eta_a,\eta_\alpha, u_j, y^{b}, y^{\beta})\quad \text{on} \quad\sT^\ast\sT^\ast M, \end{equation}
i.e., for $\psi\in\sT^\ast\sT^\ast M$ we can write $\psi=u_j(\psi)\xd q^j+y^a(\psi)\xd \eta_a+y^\alpha(\psi)\xd \eta_\alpha$.
Again, we shall write $\Lambda_M^{\#}=\beta_M^{-1}$ in the new coordinates as
\begin{equation}\label{eq:13}
(\,q^i, \eta_a,\eta_\alpha, \dot q^k, \dot \eta_b, \dot\eta_\beta\,)\circ \beta_M^{-1}=
(\,q^i, \eta_a,\eta_\alpha,\, \rho^{k}_ay^a+\rho^{k}_\alpha y^\alpha,\, c^i_{bj}\eta_iy^j-\rho^i_ba_i,\, c^i_{\beta j}\eta_ix^j-\rho^i_\beta u_i ).
\end{equation}
Recall that a mechanical system is {\it hyperregular} if the Legendre map $\lambda: \sT M\rightarrow \sT^\ast M$, defined as $\lambda=\xi_{\sT M}\circ\xd L$, is a diffeomorphism. In such a case, the dynamics is the image of a Hamiltonian vector field, and can be obtained in several equivalent ways
$$D=X_H(\sT^\ast M)=\Lambda_M^\#(\xd H(\sT^\ast M))=\varepsilon_M(\xd L(\sT M)),$$
where the Lagrangian $L$ and the Hamiltonian $H$ are related by the usual Legendre transformation
$$H(\phi)=\langle\phi, \lambda^{-1}(\phi)\rangle-L(\lambda^{-1}(\phi)).$$
For not hyperregular systems the dynamics is an implicit differential equation. Formally, from the Lagrangian point of view, it looks the same, i.e., $D=\varepsilon_M(\xd L(\sT M))$, but the Hamiltonian description requires a more complicated generating object than just a function. There exist mechanical systems for which both Lagrangian and Hamiltonian descriptions are generalized in the sense that the dynamics is not given by a Lagrangian nor Hamiltonian. Several different examples are described in \cite{slowandcareful}. The examples we will be dealing with in this paper are all hyperregular.

\subsection{Tulczyjew triple for an algebroid}\label{sec:2_2}

Once we have established the relations between an algebroid structure on a vector bundle $\tau: E\rightarrow M$ and a bivector on the dual bundle $\pi: E^\ast \rightarrow  M$, we can build the Tulczyjew triple for an algebroid $E$, using maps $\Lambda^\#$, associated with the bivector, and $\varepsilon$, associated with the bracket. The appropriate diagram is the following
$$\xymatrix@C-10pt{
 & \sT^\ast E^\ast\ar[dr]\ar[ddl]\ar[rrr]^{\Lambda^\#} & & & \sT E^\ast\ar[dr]\ar[ddl] & & & \sT^\ast E\ar[lll]_\varepsilon\ar[rd]\ar[ldd] & \\
 & & E\ar[rrr]^(0.65)\rho\ar[ddl] & & & \sT M\ar[ddl] & & & E\ar[lll]_(0.65)\rho\ar[ldd]\ar@/_1pc/[ul]_{\color{red}d L} \\
 E^\ast\ar@{.}[rrr]\ar[dr]\ar@/^1pc/[uur]^{\color{red}d H} & & & E^\ast \ar[dr]\ar@/^1pc/[uur]^{\color{red}X_H} & & & E^\ast\ar@{.}[lll]\ar[rd] & & \\
 & M\ar@{.}[rrr] & & & M & & & M\ar@{.}[lll] &
}$$
Again, identities are represented as dotted lines. The details of the construction, as well as several examples, can be found in \cite{GGU2006}. The left-hand side is Hamiltonian, the right-hand side is Lagrangian, and the dynamics lives in the middle, i.e., it is a submanifold of $\sT E^\ast$ given as $D_L=\varepsilon(\xd L(E))$ or $D_H=\Lambda^\#(\xd H(E^\ast))$. The dynamics $D_H$ generated by a Hamiltonian function is always explicit, i.e., it is an image of a vector field, while the dynamics generated by a Lagrangian may be implicit. In regular cases, the dynamics can be generated both by a Lagrangian and a Hamiltonian, in non-regular cases, more sophisticated generating objects may be needed. In other words, the algebroid triple works just as the classical one; $\Lambda^\#$ and $\varepsilon$ are not isomorphisms though.

Introducing coordinates $(q^i)$ in the base and picking a basis $(f_A)$ of sections of the bundle $\tau: E\rightarrow M$ over the coordinate domain in $M$, we get the adapted coordinates $(q^i, x^A)$ in $E$, $(q^i, \eta_A)$ in $E^\ast$, the structure functions $c^A_{BD}$ such that $[f_D, f_B]=c^A_{BD}f_A$, and the matrix elements $\rho^i_A$ for the anchor. Next, we construct adapted coordinate systems in tangent and cotangent bundles of $E$ and $E^\ast$. We get then
$$
\begin{array}{lll}\hspace{10pt}
(q^i, \eta_A, \dot q^j, \dot \eta_A) & \text{in} & \sT E^\ast, \\ \hspace{10pt}
(q^i, \eta_A, u_j, y^B) & \text{in} & \sT^\ast E^\ast, \\ \hspace{10pt}
(q^i, x^A, a_j, \xi_B) & \text{in} & \sT^\ast E.
\end{array}
$$
The coordinate expressions for $\Lambda^\#$ and $\varepsilon$ are 
\begin{equation}\label{eq:19}
(q^i, \eta_A, \dot q^j, \dot\eta_B)\circ\Lambda^\#=(q^i, \eta_A, \rho^j_Ay^A, c^D_{BE}\eta_Dy^E-\rho^i_Bu_i)
\end{equation}
and
\begin{equation}\label{eq:20}
(q^i, \eta_A, \dot q^j, \dot\eta_B)\circ\varepsilon=(q^i, \xi_A, \rho^j_Ax^A, c^D_{BE}\eta_Dy^E+\rho^i_Ba_i).
\end{equation}
The Hamiltonian vector field for a Hamiltonian $H:E^\ast\rightarrow \R$ reads then
$$X_H=\rho^j_A\frac{\partial H}{\partial \eta_A}\partial_{q^i}+\left(c^D_{BE}\eta_D\frac{\partial H}{\partial \eta_E}-\rho^i_B\frac{\partial H}{\partial q^i}\right)\partial_{\eta_B}.$$
The dynamics given by a Lagrangian function $L:E\rightarrow \R$ is described by the equations
$$\eta_A=\frac{\partial L}{\partial x^A},\qquad\dot q^i=\rho^i_A x^A, \qquad \dot\eta_{B}=c^D_{BE}\frac{\partial L}{\partial x^D}x^E+\rho^i_B\frac{\partial L}{\partial q^i},$$
in which $(x^A)$ are parameters. The Euler-Lagrange equations for curves $t\mapsto (q^i(t), x^A(t))$ in $E$, which are consequences of the above dynamical equations are 
\begin{equation}\label{eq:21}\frac{\xd}{\xd t}(q^i)=\rho^i_A x^A, \qquad\frac{\xd}{\xd t}\left(\frac{\partial L}{\partial x^{B}}\right)=c^D_{BE}\frac{\partial L}{\partial x^D}x^E+\rho^i_B\frac{\partial L}{\partial q^i}
\end{equation}

Note that in definition \ref{d:1} we have insisted that the bracket is a Lie bracket, i.e., it is antisymmetric and fulfills the Jacobi identity. This last condition, however, does not play any role in deriving dynamical equations. We may as well use just an antisymmetric bracket and a corresponding bivector $\Lambda$, not necessarily a Poisson one. Such a structure is called a {\it skew algebroid}. The lack of the Jacobi identity affects the properties of dynamical equations, but not the way they are obtained. This is precisely the case of the following example of the system with nonholonomic constraints and a special type of a Lagrangian.
\medskip

\begin{example}[Skew algebroid structure for a constrained system with mechanical Lagrangian]\label{ex:1} In this example, we shall construct a skew algebroid that can be used for deriving phase equations for a constrained system with a Lagrangian of a certain type. We assume that $M$ is a manifold with a Riemannian metric $g$. The subbundle $C\subset \sT M$ will play the role of linear constraints for a system described by the Lagrangian
\begin{equation}\label{eq:2} L:\sT M\ni v\longmapsto \frac{m}{2}g(v,v)-V(\tau_M(v))\in\R,\end{equation}
or, since such a Lagrangian is hyperregular, the Hamiltonian
\begin{equation}\label{eq:8} H:\sT^\ast M\ni p\longmapsto \frac{1}{2m}g(p,p)+V(\pi_M(p))\in\R.\end{equation}
Such a Lagrangian is often called in the literature {\it a mechanical Lagrangian}.

The metric gives us the possibility to define the orthogonal complement $C^\perp$ of $C$. We have then $\sT M=C\oplus_MC^\perp$, the inclusion map $\imath: C\hookrightarrow \sT M$, and the projection map $\pi: \sT M\rightarrow C$. With the orthogonal complement $C^\perp$ chosen, the cotangent bundle can also be written as $\sT^\ast M=C^\circ\oplus_M C^\ast$, where $C^\circ$ is the annihilator of $C$ and $C^\ast$ its orthogonal complement, naturally identified with the dual of $C$.

In section \ref{sec:2_1}, we introduced coordinates in on $\sT M$ and $\sT^\ast M$ adapted to constraints, replacing the usual $(q^i, \dot q^j)$ and $(q^i, p_j)$. In the presence of a metric, the basis $(f_i)$ of sections of $\sT M$ and the dual basis  $(\phi^j)$ of sections of $\sT^\ast M$ can be chosen in such a way that the mixed components of the metric vanish, $g_{a\alpha}=g_{\alpha a}=0$. It means that $C_q=\mathsf{Span}\{f_1(q),\ldots f_k(q)\}$,  $C^\perp_q=\mathsf{Span}\{f_{k+1}(q),\ldots f_n(q)\}$, and similarly, $C^\ast_q=\mathsf{Span}\{\phi^1(q),\ldots \phi^k(q)\}$,  $C^\circ_q=\mathsf{Span}\{\phi^{k+1}(q),\ldots \phi^n(q)\}$. Coordinates associated to $(f_i)$ will be as previously denoted by $(x^i)$, sometimes split into groups $(x^a, x^\alpha)$ with $a\in\{1,\ldots, k\}$, $\alpha\in\{k+1,\ldots, n\}$. The coordinates associated to $(\phi^i)$ will be denoted by $(\eta_i)$ and split into groups $(\eta_a, \eta_\alpha)$.

We will also use coordinates in iterated tangent and cotangent bundles. We will follow conventions set as in (\ref{eq:3}, \ref{eq:17}, \ref{eq:4}), i.e., we will use
$$
\begin{array}{lll}\hspace{10pt}
(q^i, \eta_a, \eta_\alpha, \dot q^j, \dot \eta_b, \dot \eta_\beta ) & \text{in} & \sT \sT^\ast M, \\ \hspace{10pt}
(q^i, \eta_a, \eta_\alpha, u_j, y^b, y^\beta) & \text{in} & \sT^\ast \sT^\ast M, \\ \hspace{10pt}
(q^i, x^a, x^\alpha, a_j, \xi_b, \xi_\beta) & \text{in} & \sT^\ast \sT M.
\end{array}
$$
\smallskip

From the variational principle, one can obtain the constrained Euler-Lagrange equations, which in the coordinates $(q^i,x^a, x^\alpha)$ and for a general Lagrangian $L$ read
\begin{equation}\label{eq:31}
\frac{\xd}{\xd t}(q^i)=\rho^i_ax^a, \quad
\frac{\xd}{\xd t}\left(\frac{\partial L}{\partial x^b}\right)=
c^a_{bd}x^d\frac{\partial L}{\partial x^a}+c^\alpha_{bd}x^d\frac{\partial L}{\partial x^\alpha}+\rho^i_b\frac{\partial L}{\partial q^i}, \quad x^\alpha=0.
\end{equation}
Here, as previously, $\rho$ denotes the matrix of coordinate change, i.e., $\dot q^i=\rho^i_ax^a+\rho^i_\alpha x^\alpha$ and functions $c^a_{bd}$ and $c^\alpha_{bd}$ describe the Lie bracket of sections forming the basis, $[f_d,f_b]=c^a_{bd}f_d+c^\alpha_{bd}f_\alpha$. If the functions $c^\alpha_{bd}$ are not equal to zero, then the subbundle $C$ is not integrable, therefore it is not a Lie subalgebroid of the tangent bundle.
Equation (\ref{eq:31}) looks almost like a coordinate expression of the Euler-Lagrange equations on an algebroid (see (\ref{eq:21})). The only obstruction is the term $\frac{\partial L}{\partial x^\alpha}$ which depends on values of $L$ outside $C$. However, if the Lagrangian function is of the form (\ref{eq:2}), then
$$\frac{\partial L}{\partial x^\alpha}=
mg_{\alpha a}x^a+mg_{\alpha\beta}x^\beta=0,$$
because $g_{\alpha a}=0$ everywhere and $x^\beta=0$ on $C$. It is then possible to describe the dynamics of the constrained system with mechanical Lagrangian using the structure of a skew algebroid on $C$ and the Lagrangian function $L_C$ being  $L$ restricted to $C$. We have then
$$L_C(q^i, x^a)=\frac{m}{2}g_{ab}x^ax^b-V(q)$$
and Euler-Lagrange equations
$$
\frac{\xd}{\xd t}(q^i)=\rho^i_ax^a, \quad
\frac{\xd}{\xd t}\left(\frac{\partial L_C}{\partial x^b}\right)=
c^a_{bd}x^d\frac{\partial L_C}{\partial x^a}+\rho^i_b\frac{\partial L_C}{\partial q^i}.
$$
Using the special form of $L_C$, we get
\begin{equation}\label{eq:22}
\frac{\xd}{\xd t}(q^i)=\rho^i_ax^a, \quad
m\frac{\xd}{\xd t}\left(g_{ab}x^a\right)=
mc^a_{bd}x^dg_{ae}x^e+\rho^i_b\left(\frac{m}{2}\frac{\partial g_{ae}}{\partial q^i}x^ax^e-\frac{\partial V}{\partial q^i}\right).
\end{equation}
It is a particular case of a more general situation discussed in \cite{Grabowski2008}.
\smallskip

Let us examine the geometry of our system. To do this, we shall need three components. The first is the Tulczyjew map $\alpha_M$, which is one of the ways of encoding the canonical Lie algebroid structure of $\sT M$. The second ingredient is the tangent map $\sT \imath^\ast: \sT\sT^\ast M\rightarrow\sT C^\ast$. Recall that $\imath$ denotes the inclusion $\imath: C\hookrightarrow \sT M$, its dual $\imath^\ast$ is then a projection $\imath^\ast: \sT^\ast M\rightarrow C^\ast$. Finally, we construct the third ingredient: dualizing the map tangent to $\pi:\sT M\rightarrow C$, we obtain the linear relation $(\sT\pi)^\ast:\sT^\ast C\rel \sT^\ast \sT M$, which can be transformed into a map by restricting the codomain to $\sT^\ast_C\sT M$. As a result, we get a map $(\sT\pi)^\ast_C:\sT^\ast C\rightarrow \sT_C^\ast \sT M$. The composition
$$\varepsilon_C:\sT^\ast C\longrightarrow \sT C^\ast, \quad \varepsilon_C=(\sT\imath^\ast)\circ\alpha_M^{-1}\circ(\sT\pi)^\ast_C,$$
is the double vector bundle morphism defining the structure of a skew algebroid on $C$. Let us first write this map in coordinates. The coordinate expression for the inverse of the Tulczyjew isomorphism is (\ref{eq:12}). For the other ingredients, we have
$$(q^i, \eta_a)\circ\imath^\ast=(q^i,\eta_a),\qquad (q^i, \eta_a,\dot q^i, \dot \eta_a)\circ\sT\imath^\ast=(q^i,\eta_a, \dot q^i, \dot \eta_a),$$
which means that the projection $\imath^\ast$ on $C^\ast$ is just forgetting the last group of coordinates $(\eta_\alpha)$.
Elements $\varphi \in \sT^\ast\sT M$ and $\psi\in \sT^\ast C$ are in the relation $(\sT\pi)^\ast$ if
$$q^i(\varphi)=q^i(\psi),\quad x^a(\varphi)=x^a(\psi), \quad a_i(\varphi)=a_i(\psi),\quad \xi_a(\varphi)=\xi_a(\psi),\; \text{ and }\;\xi_\alpha(\varphi)=0.$$
Adding the condition $x^\alpha(\varphi)=0$ means restricting the codomain to $\sT^\ast_C\sT M$.  The map $\varepsilon_C$ in coordinates reads
$$(q^i,\eta_a,\dot q^k, \dot\eta_b)\circ\varepsilon_C=(q^i, \xi_a, \rho^k_a x^a,  c^a_{bd}x^d\xi_a+\rho^i_b a_i).$$
It looks formally like a map associated with a Lie algebroid, with the bracket of sections given by the functions $c^a_{bd}$, and the anchor $\rho$ with the matrix $\rho^i_a$. In case $C$ is not a subalgebroid, the bracket given as
$$[f_d,f_b]_C=c^a_{bd}f_a$$
does not satisfy the Jacobi identity. It is skew-symmetric, though, therefore $C\rightarrow M$ is a skew algebroid.

In terms of the bracket, this structure can be characterized in a very straightforward way: The bracket of two sections $X$, $Y$ of the constraint $C$ is given as $[X,Y]_C=[X,Y]^{\|}$ where $[\,\cdot\, ,\,\cdot\,]$ is the usual Lie bracket of vector fields  and the superscript denotes taking the orthogonal projection on $C$.

The dynamics, given by a mechanical Lagrangian of the form (\ref{eq:2}) restricted to $C$, reads then
\begin{equation}\label{eq:9}
\begin{aligned}
\eta_a &=m g_{ab}x^b, \\
\dot\eta_b & =mc^a_{bd}x^dg_{ae}x^e+\rho^i_b \left(\frac{m}{2}\frac{\partial g_{ad}}{\partial q^i}x^ax^d-\frac{\partial V}{\partial q^i}\right), \\
\dot q^i& =\rho^i_ax^a.
\end{aligned}
\end{equation}
As a consequence of the above dynamics, we get the following Euler-Lagrange equations for a curve $t\mapsto (q^i(t), x^a(t))$ in $C$:
$$
m\frac{\xd}{\xd t}\left(g_{ab}x^a\right)=mc^a_{bd}x^dg_{ae}x^e+\rho^i_b \left(\frac{m}{2}\frac{\partial g_{ad}}{\partial q^i}x^ax^d-\frac{\partial V}{\partial q^i}\right), \qquad
\dot q^i=\rho^i_ax^a,
$$
which are (\ref{eq:22}) obtained from the variational principle. Since the mechanical Lagrangian, as well as its restriction to constraints, are hyperregular, we can get rid of the parameters $(x^a)$ and write the dynamics (\ref{eq:9}) in the explicit form
\begin{align*}
\dot\eta_b & =\frac{1}{m}c^a_{bd}\eta_ag^{de}\eta_e-\rho^i_b \left(\frac{1}{2m}\frac{\partial g^{ae}}{\partial q^i}\eta_a\eta_e+\frac{\partial V}{\partial q^i}\right), \\
\dot q^i& =\rho^i_ag^{ab}\eta_b.
\end{align*}
The above equations are, actually, Hamiltonian equations for the mechanical Hamiltonian of the form (\ref{eq:8}) with respect to the  bivector field $\Lambda_C$, which in coordinates reads
$$\Lambda_C=\rho^k_b\frac{\partial}{\partial \eta_b}\wedge \frac{\partial}{\partial q^k}+\frac{1}{2}c^a_{bd}\eta_a\frac{\partial}{\partial \eta_b}\wedge \frac{\partial}{\partial \eta_d}.$$
The coordinate-free definition of $\Lambda_C$ is as follows: for $\varphi,\psi\in\sT^\ast C^\ast$ we have
$$\Lambda_C(\varphi,\psi)=\langle \varphi, \varepsilon_C\circ \mathcal{R}_C(\psi)\rangle.$$

Summarizing, using the metric we were able to define a skew algebroid structure $\varepsilon_C$ on the constraint subbundle $C$. It corresponds to the linear bivector $\Lambda_C$ on the dual $C^\ast$. The dual bundle can be considered a phase space for systems with constraints $C$ and mechanical Lagrangians. For such a system, dynamics is given by a vector field on the phase space. This vector field depends only on the Lagrangian restricted to constraints and, moreover, it is a Hamiltonian vector field for the restricted Hamiltonian defined on $C^\ast$, and obtained by the usual Legendre transformation from the restricted Lagrangian. The name {\it restricted Hamiltonian} is consistent with the fact that it is actually a full Hamiltonian restricted to $C^\ast$. The geometric picture for a constrained system with a mechanical Lagrangian or Hamiltonian is almost the same as for the unconstrained case, with one difference: Lie algebroid is replaced by a skew algebroid, or linear Poisson structure is replaced by a linear almost Poisson structure.
\end{example}
\medskip

\begin{example}[A skater]\label{ex:8}
A configuration manifold of a skate is $M=\R^2\times S^1$ with coordinates $(x,y,\varphi)$. Since linear velocity $\dot x\partial_x+\dot y\partial_y$ of an ideal skate has to be parallel to the skate itself, the system is constrained with the constraint subbundle
$$C=\langle \cos\varphi\partial_x+\sin\varphi\partial_y, \partial_\varphi\rangle.$$
\begin{center}
\begin{tikzpicture}
\filldraw[gray!20!white] (-2,-1.5) -- (2,-1.5) -- (1.5,-.5) -- (-1.5,-.5) -- (-2,-1.5);
\draw (-2,-1.5) -- (2,-1.5) -- (1.5,-.5) -- (-1.5,-.5) -- (-2,-1.5);
\filldraw[lightgray] (-1,0) arc [radius=1, start angle=180, end angle=360];
\draw[black] (-1,0) arc [radius=1, start angle=180, end angle=360];
\draw (-1,0) -- (1,0);
\draw[thick] (0,-1) -- (0,1);
\draw[gray, ->] (-.15,.7) arc [radius=.2, start angle=135, end angle=405];
\filldraw (0,-1) circle [radius=0.03];
\node[right] at (0,-1.2) {$(x,y)$};
\node[right] at (.2,.7) {$\varphi$};
\end{tikzpicture}
\end{center}
In this example, $M$ is a metric manifold with the metric
$$g=\xd x\otimes\xd x+\xd y\otimes\xd y+k^2\xd \varphi\otimes\xd \varphi$$ where $k^2=I/m$ is a constant related to the physical parameters of the skate (its mass and moment of inertia). In our calculations, we shall use the following basis of sections of $\sT M$ adapted to the constraints:
$$f_1=\cos\varphi\partial_x+\sin\varphi\partial_y,\quad f_2=\partial_\varphi, \quad f_3=-\sin\varphi\partial_x+\cos\varphi\partial_y,$$
with corresponding coordinates $(z^i)$. The constraints are given by the condition $z^3=0$. Note, that $f_3$ is orthogonal to $f_1$ and $f_2$. Coordinates of the velocity change according to the formula
\begin{equation}\label{eq:32}\left[
\begin{array}{c}
\dot x \\ \dot y\\ \dot \varphi\end{array}\right]=
\left[
\begin{array}{ccc}
\cos\varphi & 0 & -\sin\varphi \\
\sin\varphi & 0 & \cos\varphi \\
0 & 1 & 0
\end{array}\right]
\left[
\begin{array}{c}
z^1 \\ z^2 \\ z^3 \end{array}\right].
\end{equation}
We have then
$$\left[\rho^i_a\right]= \left[\begin{array}{cc}
\cos\varphi & 0  \\
\sin\varphi & 0  \\
0 & 1
\end{array}\right] \qquad \text{and}\qquad \left[\rho^i_\alpha\right]=
\left[\begin{array}{c}
 -\sin\varphi \\
 \cos\varphi \\
0
\end{array}\right],
$$
with $i\in\{1,2,3\}$, $a\in\{1,2\}$ and $\alpha=3$. Calculating the brackets of vector fields $f_i$, we get
$$[f_1,f_2]=-f_3,\quad [f_1,f_3]=0,\quad [f_2,f_3]=-f_1,$$
which shows that $C=\langle f_1, f_2\rangle$ is not integrable. If, as usual, we introduce structure functions $c^i_{jk}$ by the formula $[f_k,f_j]=c^i_{jk}f_i$, we get $c^3_{12}=1$ and $c^1_{23}=1$, while the rest of the structure functions vanish.

On $\sT^\ast C$, we have coordinates $(x,y,\varphi;\, z^1,z^2;\, p_x, p_y,p_\varphi; \xi_1, \xi_2)$, where coordinates $(p,\xi)$ are adapted coordinates on a cotangent bundle. On $C^\ast$, coordinates are $(x,y,\varphi;\, \eta_1, \eta_2)$, with coordinates $(\eta_a)$ being dual to $(z^a)$. On $\sT C^\ast$, we shall use coordinates $(x,y,\varphi;\, \eta_1, \eta_2;\, \dot x, \dot y, \dot\varphi;\, \dot \eta_1,\dot\eta_2)$, where the `dotted' coordinates are constructed as usual. The skew-algebroid structure on $C$, induced by the canonical Lie algebroid structure on $\sT M$ and the metrics reads
$$
(x,y,\varphi;\, \eta_1, \eta_2;\, \dot x, \dot y, \dot\varphi;\, \dot \eta_1,\dot\eta_2)\circ\varepsilon_C=
(x,y,\varphi;\, \xi_1, \xi_2;\, z^1\cos\varphi, z^1\sin\varphi, z^2;\, a_x\cos\varphi+a_y\sin\varphi, a_\varphi),
$$
while the corresponding almost-Poisson bivector on $C^\ast$ in coordinates is
$$\Lambda_C=\cos\varphi\partial_{\eta_1}\wedge\partial_x+\sin\varphi\partial_{\eta_1}\wedge\partial_y+
\partial_{\eta_2}\wedge\partial_{\varphi}.$$
The free skater is described by a free Lagrangian defined on constraints,
$$L_0(x,y,\varphi;\, z^1,z^2)=\frac{m}{2}\left((z^1)^2+k^2(z^2)^2\right),$$
while for the skater on the slope we have
$$L_1(x,y,\varphi;\, z^1,z^2)=L_0(x,y,\varphi;\, z^1,z^2)-\lambda x.$$
 The corresponding Hamiltonians are, of course,
$$H_0(x,y,\varphi;\, \eta_1,\eta_2)=\frac{1}{2m}\left((\eta_1)^2+\frac{1}{k^2}(\eta_2)^2\right), \qquad
H_1(x,y,\varphi;\, \eta_1,\eta_2)=H_0(x,y,\varphi;\, \eta_1,\eta_2)+\lambda x.$$

The phase equations can be obtained both from Lagrangians and Hamiltonians. The dynamics is a subset of $\sT C^\ast$ given by
$$\mathcal{D}=\varepsilon_C(\xd L(C))=\Lambda_C^\#(\xd H(C^\ast)).$$

In the free case, we get
\begin{equation}\label{eq:34}
\begin{array}{lll}\vspace{10pt}
\displaystyle\dot x = \frac{\cos\varphi}{m} \eta_1, & \quad & \dot\eta_1 = 0, \\ \vspace{10pt}
\displaystyle\dot y = \frac{\sin\varphi}{m} \eta_1, & \quad & \dot\eta_2 = 0, \\
\displaystyle\dot \varphi = \frac{1}{mk^2}\eta_2.
\end{array}
\end{equation}
\smallskip

\noindent The above equations can be easily solved. What we obtain is
$$\eta_2(t)=const=mk^2\omega_0, \quad\text{and then}\quad \varphi(t)=\varphi_0+\omega_0 t.$$
Similarly,
$$\eta_1(t)=const=mv_0.$$
For $x$ and $y$, we get
$$\begin{array}{ll}\vspace{5pt}
\displaystyle x(t)=\frac{v_0}{\omega_0}\left(\sin(\varphi_0+\omega_0 t)-\sin(\varphi_0)\right)+x_0, \\
\displaystyle y(t)=-\frac{v_0}{\omega_0}\left({\cos}(\varphi_0+\omega_0 t)-\cos(\varphi_0)\right)+y_0.
\end{array}$$
As expected, the free skater, starting from $(x_0, y_0)$ in the direction of $\varphi_0$, follows the circle with the radius depending on the initial conditions $v_0/\omega_0$.
\medskip

For the skater on a slope with gravitational force, we get
\begin{equation}\label{eq:33}
\begin{array}{lll}\vspace{10pt}
\displaystyle\dot x = \frac{\cos\varphi}{m} \eta_1, & \quad & \dot\eta_1 = -\lambda\cos\varphi, \\ \vspace{10pt}
\displaystyle\dot y = \frac{\sin\varphi}{m} \eta_1, & \quad & \dot\eta_2 = 0, \\
\displaystyle\dot \varphi = \frac{1}{mk^2}\eta_2.
\end{array}
\end{equation}
\smallskip

\noindent The equations for $\eta_2$ and $\varphi$ are as before, therefore we get
$$\eta_2(t)=mk^2\omega_0, \quad \varphi(t)=\varphi_0+\omega_0 t.$$
We solve the equation for $\eta_1$,
$$\eta_1(t)=-\frac{\lambda}{\omega_0}\sin(\varphi_0+\omega_0 t)+v_0m,$$
and integrate the equations for $x$ and $y$,
$$
\begin{array}{l}\vspace{10pt}
\displaystyle x(t)=\frac{\lambda}{4m\omega_0^2}\cos(2\varphi_0+2\omega_0 t)+\frac{v_0}{\omega_0} \sin(\varphi_0+\omega_0 t)+x_0 \\
\displaystyle y(t)=-\frac{\lambda}{2m\omega_0}t+\frac{\lambda}{4m\omega_0^2}\sin(2\varphi_0+2\omega_0 t)-\frac{v_0}{\omega_0}\cos(\varphi_0+\omega_0 t)+y_0.
\end{array}
$$
In the above equations, we did not follow the rules that $x_0, y_0, v_0$ denote initial positions and velocity. With such parametrization, formulae become difficult to read. Below we can see an example of the trajectory of the point of contact of the skate with the ice. As we can see, the skate does not slide from the slope, but rather follows the cycloid motion in the direction perpendicular to the slope, which is a well-known phenomenon.
\begin{center}
\includegraphics[scale=0.5]{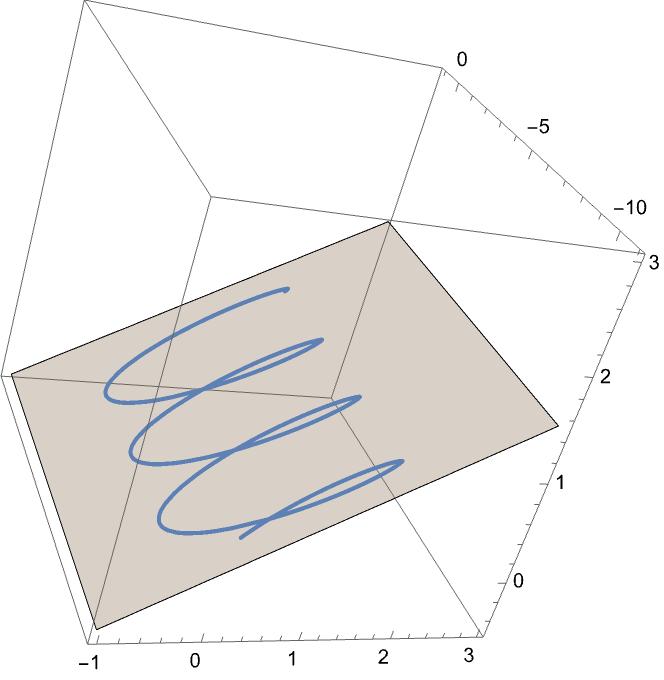}
\end{center}
\end{example}

\begin{example}[Almost Poisson structure for constrained nonholonomic systems with gyroscopic terms]\label{ex:2} Following \cite{Garcia2023}, we shall now discuss a constrained mechanical system with a Lagrangian containing a magnetic or gyroscopic term, more precisely a Lagrangian of the form
\begin{equation}\label{eq:10}
L:\sT M\longrightarrow\R,\quad L(v)=\frac{m}{2}g(v,v)-\langle A(\tau_M(v)),v\rangle-V(\tau_M(v)),
\end{equation}
where $A$ is a one-form on $M$ representing, e.g., a magnetic potential. Such Lagrangian is hyperregular and corresponds to the Hamiltonian
\begin{equation}\label{eq:10a}
H:\sT^\ast M\longrightarrow\R,\quad H(p)=\frac{1}{2m}g(p-A(\pi_M(p)),p-A(\pi_M(p)))+V(\pi_M(p)).
\end{equation}
The constraint subbundle will be, as usual, denoted by $C$, with the dual $C^\ast$. The dynamics of this system can be described in Hamiltonian-like form, using an almost-Poisson bracket on the phase space $C^\ast$. The only problem is that, to construct this bracket, one has to use the magnetic potential that appears in the Lagrangian and Hamiltonian. Consequently, the structure is not universal; it depends on the functions generating the dynamics.
\smallskip

We shall discuss the Hamiltonian formulation of the dynamics first. Recall that, due to the metric, we have the splitting $\sT^\ast M=C^\circ\oplus C^\ast$. We can therefore write $A$ as a sum $A(q)=A(q)^\perp+A(q)^\parallel$ where $A(q)^\perp\in C^\circ$ and
$A(q)^\parallel\in C^\ast$.  Let $\chi: C^\ast\rightarrow \sT^\ast M$ be the embedding $\chi(p)=p+A^\perp(\pi_M(p))$, i.e., we identify $C^\ast$ with an affine subbundle $\mathcal{E}$ of $\sT^\ast M$ being $C^\ast$ translated by the perpendicular part of $A$.
In coordinates, $\chi$ reads
$$(q^i, \eta_a, \eta_\alpha)\circ\chi=(q^i, \eta_a, A_\alpha(q)).$$
The tangent map $\sT\chi$ is also an embedding of $\sT C^\ast$ into $\sT\sT^\ast M$, or, more precisely, into $\sT_\mathcal{E}\sT^\ast M$, i.e. the image of $\sT\chi$ projects by $\tau_{\sT^\ast M}$ on $\mathcal{E}$. Taking the dual $\sT^\ast\chi$, we obtain a linear relation between $\sT^\ast C^\ast$ and $\sT^\ast\sT^\ast M$. Taking into account only elements of $\sT^\ast_\mathcal{E}\sT^\ast M$, this relation becomes a projection from $\sT^\ast_\mathcal{E}\sT^\ast M$ to $\sT^\ast C^\ast$. If we restrict the domain even more, requiring that element of $\sT^\ast_\mathcal{E}\sT^\ast M$ projects by $\xi_M$ on $C$, we obtain the map from 
$$\pi_{\sT^\ast M}^{-1}(\mathcal{E})\cap\xi_M^{-1}(C)\subset \sT^\ast\sT^\ast M\quad\text{to}\quad \sT^\ast C^\ast,$$ 
which is invertible. Taking the inverse, we can embed $\sT^\ast C^\ast$ into $\sT^\ast\sT^\ast M$. This embedding will be denoted by $(\sT^\ast\chi)_C$. In coordinates, we get
$$ (q^i, \eta_a,\eta_\alpha, u_j, y^b, y^{\beta})\circ (\sT^\ast\chi)_C=(q^i, \eta_a, A_{\alpha}(q), u_j, y^b, 0).$$
Now, we can compose $(\sT^\ast\chi)_C$ with $\beta_M$ and $\sT\imath^\ast$, and define $\beta_A$:
$$\beta_A:\sT^\ast C^\ast\rightarrow\sT C^\ast, \qquad \beta_A=\sT \imath^\ast\circ\beta_M^{-1}\circ(\sT^\ast\chi)_C.$$
In coordinates, we have
$$(q^i, \eta_a, \dot q^j, \dot \eta_b)\circ\beta_A=(q^i, \eta_a, \rho^k_by^b, c^a_{bd}\eta_yx^d+c^\alpha_{bd}A_\alpha y^d-\rho^l_b u_l).$$
Note that, this time, the map $\beta_A$ is not a double vector bundle morphism. It is a morphism of vector bundles over $C^\ast$, but as a map between bundles over $\sT M$ and $C$ it is affine. This means that $\beta_M$ corresponds to a certain bivector, but this bivector is not linear. Consequently, it does not define an algebroid structure on $C$.

The dynamics of the system is given as $\beta_A(\xd H_C(C^\ast))\subset\sT C^\ast$, with $H_C(p)=H(\chi(p))$,
and described in coordinates as
\begin{equation}\label{eq:11}
\begin{aligned}
\dot q^k& =\rho^k_b\frac{\partial H_C}{\partial \eta_b}, \\
\dot \eta_b &= c^a_{bd}\eta_a\frac{\partial H_C}{\partial \eta_d}+c^\alpha_{bd}A_\alpha \frac{\partial H_C}{\partial \eta_d}-\rho^l_b \frac{\partial H_C}{\partial q^l},
\end{aligned}
\end{equation}
with $$H_C(q^i,\eta_a)=\frac{1}{2m}g^{ab}(\eta_a-A_a(q))(\eta_b-A_b(q))+V(q).$$
Equations (\ref{eq:11}) are explicit, which means that they are given by a vector field on the phase space.

As we have mentioned before, the map $\beta_A$ is associated to a bivector field $\Lambda_A$ on $C^\ast$, defined as
$$\Lambda_A(\varphi,\psi)=\langle \varphi, \beta_A(\psi)\rangle.$$
In coordinates $(q^i,\eta_a)$, it reads
$$\Lambda_A(q,\eta)= \rho(q)^k_b\frac{\partial }{\partial \eta_b}\wedge \frac{\partial }{\partial q^k}+
\frac{1}{2}\left[c^a_{bd}(q)\eta_a+c^\alpha_{bd}(q)A_\alpha(q)\right]\frac{\partial }{\partial \eta_b}\wedge \frac{\partial }{\partial \eta_d}.$$
This is precisely the almost-Poisson structure defined in \cite{Garcia2023}.

Since the Hamiltonian $H_C$ is hyperregular, it is easy to construct a Lagrangian description of our mechanical system with magnetic force and linear constraints. Using the canonical double vector bundle isomorphism $\mathcal{R}_C:\sT^\ast C^\ast\rightarrow \sT^\ast C$, we define the map $\varepsilon_A$ by the formula
$$\varepsilon_A: \sT^\ast C\rightarrow\sT C^\ast, \qquad \varepsilon_A=\beta_A\circ \mathcal{R}_C^{-1}.$$
Since $\beta_A$ is not a double vector bundle morphism, it is also the case of $\varepsilon_A$. It means that the Lagrangian formulation is not given by an algebroid bracket on $C$, but rather by some affine structure. The Lagrangian function $L_C$ defined on constraints can be obtained from the Hamiltonian $H_C$ by the usual formula
$$L_C(v)=\langle \lambda_C(v), v\rangle -H_C(\lambda_C(v)),$$
where $\lambda_C: C\rightarrow C^\ast$ is given by the metric restricted to $C$ and the parallel part of the magnetic potential, i.e., $\lambda_C(v)=mg(v,\cdot)-A^{\|}$. Function $L_C$ is equal to the full Lagrangian $L$ restricted to the constraint $C$. One can verify that
$$\beta_A(\xd H_C(C^\ast))=\varepsilon_A(\xd L_C(C)).$$

Summarizing, using the metric and one-form $A$, we were able to define an affine almost-Poisson bivector $\Lambda_A$ on $C^\ast$ and its morphism representation $\beta_A$. The affine character of both $\Lambda_A$ and $\beta_A$ comes from the fact that we identify $C^\ast$ with the affine subbundle $\mathcal{E}$ of $\sT^\ast M$. The canonical isomorphism $\mathcal{R}_C$ provides us with the map $\varepsilon_A$ and a Lagrangian description of the system. Lagrangian $L_C$ is defined on constraint and given by the restriction to constraint of the full Lagrangian on $\sT M$, while Hamiltonian is defined on the phase space $C^\ast$ and defined by the restriction of the full Hamiltonian to the affine subbundle $\mathcal{E}\subset \sT^\ast M$.
\end{example}

\section{Dirac structures in nonholonomic mechanics}\label{sec:3}

In section \ref{sec:2}, we have used a version of the Tulczyjew triple based on the structure of a skew algebroid to generate phase equations of a system with nonholonomic constraints and a special type of Lagrangian. Metrics on the manifold of positions were an important part of the structure. The modified, affine version of the same type of structure was used for a system with a magnetic-like force. The problem with the latter was that the geometric structure we used to derive phase equations depended not only on the metric but also on the magnetic potential present in the Lagrangian of the system. Part of the magnetic potential entered the almost Poisson bivector field. In the following, we propose a solution to both problems: we define a geometric structure called, a {\it Dirac algebroid}, that depends on constraints, but can be used to derive phase equations for any type of a Lagrangian or Hamiltonian -- no metric nor other specific data needed to write down a generating function is used in the construction. We test our proposal on mechanical Hamiltonians and Hamiltonians with magnetic terms.
\medskip

Let us recall the necessary definitions first. For a manifold $N$, we consider the Whitney sum of its cotangent and tangent bundle $\sT^\ast N\oplus_N\sT N$, together with the symmetric two-form with signature $(\dim N,\dim N)$,
$$(\alpha_1+X_1|\alpha_2+X_2)=\langle\alpha_1, X_2\rangle +\langle\alpha_2, X_1\rangle,$$
and the bracket of sections
$$[\![\alpha_1+X_1,\alpha_2+X_2]\!]=[X_1,X_2]+\mathcal{L}_{X_1}\alpha_2-\imath_{X_2}\xd \alpha_1,$$
called the Dorfman bracket. {\it An almost Dirac structure} $\mathcal D$ on $N$ is a subbundle of $\sT^\ast N\oplus_N\sT N$ maximally isotropic with respect to the symmetric two-form. If it is additionally closed with respect to the bracket, it is called a {\it Dirac structure}. Canonical examples of an almost Dirac structure are provided by two-forms and bivector fields. More precisely, if $\omega$ is a two-form on $N$, then its graph, defined as $\mathcal{D}_\omega=\{(\imath_v\omega, v): v\in\sT N \}\subset \sT^\ast N\oplus_N\sT N$, is an almost Dirac structure on $N$. When $\omega$ is closed, its graph is a Dirac structure. Similarly, for a bivector field $\Lambda$, the graph $\mathcal{D}_{\Lambda}=\{(\varphi, \imath_\varphi\Lambda):\; \varphi\in \sT^\ast N\}$ is an almost Dirac structure, or a Dirac structure in the case when $\Lambda$ is a Poisson tensor. For a regular distribution $\Delta\subset\sT N$ we can define an almost Dirac structure as $\mathcal{D}_{\Delta}=\Delta^\circ\oplus_N\Delta$,  where $\Delta^\circ$ is the annihilator of $\Delta$. For an integrable distribution, this becomes a Dirac structure.
\medskip

In \cite{GRABOWSKA2011} we have defined the concept of a {\it Dirac algebroid} on a vector bundle $\tau:E\rightarrow M$ as a linear almost Dirac structure on $E^\ast$, i.e., an almost Dirac structure on $E^\ast$ and, simultaneously, a double vector subbundle $\mathcal{D}$ of the double vector bundle $\sT^\ast E^\ast\oplus_{E^\ast}\sT E^\ast$. An example of such a structure is provided by a skew algebroid. Recall that a skew algebroid on the vector bundle $E\rightarrow M$  can be represented by a linear bivector field on $E^\ast$. Due to the linearity condition, the graph $\mathcal{D}_\Lambda$ is a double vector subbundle. Below we present the diagrams of the double vector bundles $\sT^\ast E^\ast\oplus_{E^\ast}\sT E^\ast$ and $\mathcal{D}_\Lambda$, with $\rho$ denoting the anchor of a skew algebroid given by $\Lambda$.
\begin{equation}\label{eq:23}
\xymatrix@R+5pt@C-5pt{
& \sT^\ast E^\ast\oplus_{E^\ast}\sT E^\ast\ar[dl]\ar[dr] & \\
E^\ast\ar[dr] & \sT^\ast M\oplus_M E^\ast\ar[d]\ar@{ (->}[u]  & E\oplus_M\sT M,\ar[dl] \\
& M &
}\qquad
\xymatrix@R+5pt{
& \mathcal{D}_\Lambda\ar[dl]_{\tau_l}\ar[dr]^{\tau_r}  & \\
E^\ast\ar[dr] & \mathrm{graph}(\rho)^\circ\ar[d]\ar@{ (->}[u]  & \mathrm{graph}(\rho).\ar[dl] \\
& M &
}
\end{equation}
With $\mathrm{graph}(\rho)$ we denoted the subbundle 
$$E\oplus_M \sT M\supset\mathrm{graph}(\rho)=\{(u,\rho(u)):\; u\in E\}.$$ 
Its (transposed) annihilator $\mathrm{graph}(\rho)^\circ$ is then a subbundle of $\sT^\ast M\oplus E^\ast$. An even more specific example, in the case $E=\sT M$, is the graph of the map $\beta_M$ associated to the canonical symplectic structure $\omega_M$ of $\sT^\ast M$.

Dirac algebroids can be used in Geometric Mechanics as a tool to generate phase equations out of Lagrangian or Hamiltonian in the presence of linear nonholonomic constraints. We treat a subbundle $\mathcal{D}$ of $\sT^\ast E^\ast\oplus_{E^\ast}\sT E^\ast$ as a relation from $\sT^\ast E^\ast$ to $\sT E^\ast$. The dynamics $D_H$ is then composed of all vectors tangent to $E^\ast$ that are in relation $\mathcal{D}$ with elements of $\xd H(E^\ast)$. Using the isomorphism $\mathcal{R}_E$, we can define an appropriate relation $\varepsilon_{\mathcal D}$ between $\sT^\ast E$ and $\sT E^\ast$, and generate the dynamics from a Lagrangian: it is composed of all vectors in relation $\varepsilon_{\mathcal D}$ with elements of $\xd L(E)$. The Tulczyjew triple defined by a Dirac algebroid is the following diagram composed of linear relations:
$$\xymatrix@C+20pt{
\sT^\ast E^\ast \ar@{-|>}[r]^{\mathcal{D}} &\; \sT E^\ast \;& \sT^\ast E \ar@{-|>}[l]_{\varepsilon_\mathcal{D}}
}
$$

The following example shows an application of a Dirac algebroid in nonholonomic mechanics.

\begin{example}[Dirac algebroid induced by linear constraints]\label{ex:7}
In this example, we follow the procedure described in \cite{GRABOWSKA2011} for the construction of a Dirac algebroid induced by a skew algebroid structure on a vector bundle $E\rightarrow M$ and linear constraints given by a linear subbundle $C$ of $E$. We will work with $E=\sT M$ and $C$ being a possibly nonintegrable distribution on $M$. It means, in principle, that there are no constraints on positions. Our initial Dirac algebroid $\mathcal{D}_{M}$ is the graph of the map $\beta_M:\sT\sT^\ast M\rightarrow \sT^\ast\sT^\ast M$. Since $\beta_M$ is invertible, we can also consider the graph of $\beta_M^{-1}$ and parameterize $\mathcal{D}_{M}$ by elements of $\sT^\ast\sT^\ast M$, which is more convenient for the intended application. We have then
$$\mathcal{D}_{M}=\{(\varphi, \beta_M^{-1}(\varphi)):\;\; \varphi\in\sT^\ast\sT^\ast M\}.$$
Using coordinates
$$(q^i,p_j,a_k,b^l) \quad\text{in}\quad \sT^\ast\sT^\ast M$$
and
$$(q^i,p_j,\dot q^k,\dot p_l) \quad\text{in}\quad \sT\sT^\ast M,$$
we can compile coordinates
$$(q^i,p_j;\,a_k,b^l;\; \dot q^s,\dot p_r)\quad\text{in}\quad \sT^\ast\sT^\ast M\oplus_{\sT^\ast M}\sT\sT^\ast M,$$
where the two first two groups of coordinates refer to $\sT^\ast\sT^\ast M$ and the first and the third group to $\sT\sT^\ast M$. The double vector structure is
$$
\xymatrix@R+5pt@C-5pt{
& \sT^\ast \sT^\ast M\oplus_{\sT^\ast M}\sT \sT^\ast M\ar[dl]\ar[dr] & \\
\sT^\ast M\ar[dr] & \sT^\ast M\oplus_M \sT^\ast M\ar[d]\ar@{ (->}[u]  & \sT M\oplus_M\sT M,\ar[dl] \\
& M &
}\qquad
\xymatrix@R+5pt{
& (q^i,p_j;\,a_k,b^l;\; \dot q^s,\dot p_r) \ar[dl]\ar[dr] & \\
(q^i,p_j)\ar[dr] & (q^i, a_k, \dot p_r)\ar[d]\ar@{ (->}[u]  & (q^i, b^l,\dot q^s).\ar[dl] \\
& (q^i) &
}
$$
The Dirac algebroid $\mathcal{D}_{M}$ reads then
$$\mathcal{D}_{M}=\{(q^i,p_j;\,a_k,b^l;\; b^s,-a_r)\}.$$
It is easy to see that $\mathcal{D}_{M}$ projects onto the whole $\sT^\ast M$ and on the graph of the identity map in $\sT M\oplus_M\sT M$. As a double vector bundle, it is
$$
\xymatrix@R+5pt@C-5pt{
& \mathcal{D}_M\ar[dl]\ar[dr] & \\
\sT^\ast M\ar[dr] & \{(a,-a):\,a\in\sT^\ast M\}\ar[d]\ar@{ (->}[u]  & \{(v,v):\, v\in\sT M\},\ar[dl] \\
& M &
}
$$
The procedure of adapting the above Dirac algebroid to the constraints consists of restricting the projection on $\sT M\oplus_M\sT M$ to the graph of the identity map on constraints $\Delta=\{(v,v):\;\; v\in C\}\subset \sT M\oplus_M\sT M$, and adding to the core of the restricted subbundle the annihilator $\Delta^\circ\subset \sT^\ast M\oplus_M\sT^\ast M$. It is easy to establish that $\Delta^\circ=\{(\phi,\psi):\; \phi+\psi\in C^\circ\}\subset \sT^\ast M\oplus_M\sT^\ast M.$

The most convenient way of working with the induced Dirac algebroid $\mathcal{D}_{C}$ is to use coordinates adapted to constraints as in (\ref{eq:3}) and (\ref{eq:4}). In these coordinates, the Dirac algebroid $\mathcal{D}_M$ reads
$$\{(q^i, \eta_a, \eta_\alpha;\; a_j, x^b, x^\beta;\; \rho^k_lx^l, c^i_{jk}\eta_ix^k-\rho^l_ja_l )\}$$
and for the Dirac algebroid $\mathcal{D}_C$ we obtain
$$\{(q^i, \eta_a, \eta_\alpha;\; a_j, x^b, 0;\; \rho^k_bx^b, c^i_{bd}\eta_ix^d-\rho^l_ba_l, \dot\eta_\alpha)\},$$
which means, in particular, that $x^\alpha=0$. Coordinates $q^i$, $\eta_a$, $\eta_\alpha$, $a_j$, $x^b$ and $\dot\eta_{\beta}$ are free parameters.

We can now use the Dirac algebroid $\mathcal{D}_C$ as the graph of a linear relation from $\sT^\ast \sT^\ast M$ to $\sT\sT^\ast M$. For a given Hamiltonian $H:\sT^\ast M\rightarrow \R$, we define the dynamics as a set $\mathcal{P}_H$ of all elements of $\sT\sT^\ast M$ that are in relation given by $\mathcal{D}_C$ with elements $\xd H(\sT^\ast M)$. For a general Hamiltonian, we obtain in coordinates the following set of equations
\begin{align}
&\frac{\partial H}{\partial \eta_\alpha}=0, \label{eq:5}\\
&\dot q^i =\rho^i_a\frac{\partial H}{\partial \eta_a}, \label{eq:6}\\
&\dot \eta_b  =c^a_{bd}\eta_a\frac{\partial H}{\partial \eta_d}+c^\alpha_{bd}\eta_\alpha\frac{\partial H}{\partial \eta_d}-\rho^l_b \frac{\partial H}{\partial q^l}. \label{eq:7}
\end{align}
Equations (\ref{eq:6}) and (\ref{eq:7}) resemble Hamilton equations; we have to notice, however, that the coordinates $\eta_\alpha$ are present in equation (\ref{eq:7}), moreover, there is no condition on $\dot\eta_\alpha$. As a result of equation (\ref{eq:5}), it may happen that the dynamics does not project on the whole $\sT^\ast M$, therefore we may have to apply some integrability conditions.
\end{example}

Let us now examine in detail two classes of Hamiltonians. First, we shall assume that our system is described by a mechanical Hamiltonian, i.e., the sum of the kinetic energy given by a metric on $M$ and a potential energy depending only on positions. This case was discussed in \cite{Grabowski2008} by means of nonholonomic brackets. Then, we shall pass to Hamiltonians with magnetic-like terms, that were described in \cite{Garcia2023}, in terms of an almost Poisson bracket. In both cases, we can determine integrability conditions and compare our results with those present in the literature.

\begin{example}[The case of mechanical Hamiltonian]\label{ex:2a} As in example \ref{ex:1}, $M$ is a metric manifold with metric $g$. The cotangent bundle $\sT^\ast M$ can, therefore, be written as the Whitney sum of $C^\circ$ and its orthogonal complement, naturally identified with $C^\ast$. We have then
$$\sT^\ast M=C^\ast\oplus_M C^\circ \quad\text{and}\quad \sT \sT^\ast M=\sT C^\ast\oplus_{\sT M}\sT C^\circ.$$
Using coordinates adapted to the splitting $\sT^\ast M=C^\ast\oplus_M C^\circ$ as in example \ref{ex:1}, we can write mechanical Hamiltonian in the form
$$H(q^i, \eta_a, \eta_\alpha)=\frac{1}{2m}g^{ab}\eta_a\eta_b+\frac{1}{2m}g^{\alpha\beta}\eta_\alpha\eta_\beta+V(q^i).$$
Condition (\ref{eq:5}) now reads
$$g^{\alpha\beta}\eta_\beta=0,$$
therefore, since $g$ is nondegenerate, $\eta_\beta=0$, and the dynamics projects on $C^\ast$. The integrability condition is now $\dot\eta_\alpha=0$. Geometrically, it means that the integrable part of the phase equation is $\mathcal{P}_H\cap\sT C^\ast$
$$ \mathcal{P}_H\cap\sT C^\ast=\left\{(q^i, \eta_a, 0,\dot q^j,\dot\eta_b, 0):\;\;
\dot q^i =\frac{1}{m}\rho^i_ag^{ab}\eta_b, \;
\dot \eta_b  =c^a_{bd}\frac{1}{m}\eta_ag^{de}\eta_e-\rho^l_b\left(\frac{1}{2m}\frac{\partial g^{de}}{\partial q^l}\eta_d\eta_e+ \frac{\partial V}{\partial q^l}\right)
\right\}. $$
Here we treat $\sT C^\ast$ as a submanifold in $\sT\sT^\ast M$. Formally, the same result we obtain using a nonholonomic bracket on $C^\ast$ and restricted Hamiltonian $H_C=H_{|C^\ast}$.
\end{example}

\begin{example}[The case of Hamiltonian with a magnetic term] We work, again, with metric manifold $(M, g)$. Let $A:M\rightarrow \sT^\ast M$ be a one-form that can, for example, represent a magnetic potential. The Hamiltonian is of the form
$$H(q^i, \eta_a,\eta_\alpha)=\frac{1}{2m}g^{ab}(\eta_a-A_a)(\eta_b-A_b)+\frac{1}{2m}g^{\alpha\beta}(\eta_\alpha-A_
\alpha)(\eta_\beta-A_\beta)+V(q^i).$$
Again, we examine condition (\ref{eq:5}), which now reads
$$g^{\alpha\beta}(\eta_\beta-A_\beta)=0,$$
therefore $\eta_\beta=A_\beta$. It means that the dynamics $\mathcal{P}_H$ projects onto the affine subbundle $\mathcal{E}$ of $\sT^\ast M$ such that $\mathcal{E}_q=C^\ast_q+A(q)^\perp$, where $A(q)^\perp$ is the component of $A(q)$ belonging to $C^\circ$. The integrability condition in coordinates reads
$$\dot\eta_\beta=\frac{\partial A_\beta}{\partial q^i}\dot q^i.$$
The integrable part of the dynamics reads
\begin{multline*} \mathcal{P}_H\cap\sT \mathcal{E}=\left\{(q^i, \eta_a, A_\alpha(q),\dot q^j,\dot\eta_b,\frac{\partial A_\beta}{\partial q^i}\dot q^i):\;\;
\dot q^i =\frac{1}{m}\rho^i_ag^{ab}\left(\eta_b-A_b\right), \right. \\
\left.\dot \eta_b  =\frac{1}{m}c^a_{bd}\eta_ag^{de}(\eta_e-A_e)-\frac{1}{m}c^\alpha_{bd}A_\alpha g^{de}(\eta_e-A_e)
-\rho^l_b\left(\frac{1}{2m}\frac{\partial}{\partial q^l}\left[g^{de}(\eta_d-A_d)(\eta_e-A_e)\right]+ \frac{\partial V}{\partial q^l}\right)
\right\}.
\end{multline*}
The integrable part of the dynamics depends only on the restriction $H_C=H_{|\mathcal{E}}$. Since there is a bijection between $C^\ast$ and $\mathcal{E}$, we can map the equation into $\sT C^\ast$. This way we obtain the same equation as in \cite{Garcia2023}. The difference is that, to define the effective phase space which is $\mathcal{E}$ and generate phase equations on it, we have used the structure of the Dirac algebroid that depends only on constraints and not on the Hamiltonian itself, as it is the case with the almost Poisson bracket introduced in \cite{Garcia2023}.
\end{example}

\begin{example}[A skater by means of the Dirac algebroid]
We come back to the example of a skate on an ice rink. We shall discuss two cases: a mechanical system with gravitational force and a system with magnetic potential. In both cases, we shall use the same Dirac structure defined by the constraints. Let us recall that the configuration manifold for a skate is $M=\R^2\times S^1$, with constraints given in natural coordinates $(x,y,\varphi)$ by
$$C=\langle \cos\varphi\partial_x+\sin\varphi\partial_y, \partial_\varphi\rangle.$$
Following the construction described in Example \ref{ex:7}, we start with the canonical Dirac structure being the graph of the symplectic form on $\sT^\ast M$. In the coordinates adapted to constraints and described in \ref{ex:8}, we have
$$
\begin{array}{lcl}\vspace{5pt}
\dot x= z^1\cos\varphi -z^3\sin\varphi,  &\quad &\dot\eta_1= \eta_3z^2-p_x\cos\varphi-p_y\sin\varphi,  \\ \vspace{5pt}
\dot y= z^1\sin\varphi +z^3\cos\varphi, &\quad & \dot\eta_2= -\eta_3z^1+\eta_1z^3-p_\varphi,\\ \vspace{5pt}
\dot\varphi= z^2, &\quad &\dot\eta_3= -\eta_1z^2+p_x\sin\varphi -p_y\cos\varphi,
\end{array}
$$
where $(x,y,\varphi, \eta_1, \eta_2, \eta_3)$ are coordinates in on $\sT^\ast M$, $(x,y,\varphi, \eta_1, \eta_2, \eta_3;\; p_x, p_y, p_\varphi, z^1, z^2, z^3)$ are coordinates on $\sT^\ast\sT^\ast M$ and
$(x,y,\varphi, \eta_1, \eta_2, \eta_3;\;  \dot x, \dot y,\dot\varphi, \dot\eta_1, \dot\eta_2, \dot\eta_3)$ are coordinates in on $\sT\sT^\ast M$.

The constraint is given by the condition $z^3=0$, which leads to equations $\dot x= z^1\cos\varphi$ and $\dot y= z^1\sin\varphi$, while $\dot\eta_3$ becomes arbitrary. The Dirac structure $\mathcal{D}_C$ is then given by the conditions
$$
\begin{array}{lcl}\vspace{5pt}
\dot x= z^1\cos\varphi,   &\quad &\dot\eta_1= \eta_3z^2-p_x\cos\varphi-p_y\sin\varphi,  \\ \vspace{5pt}
\dot y= z^1\sin\varphi,  &\quad & \dot\eta_2= -\eta_3z^1-p_\varphi,\\ \vspace{5pt}
\dot\varphi= z^2, &\quad & z^3=0,
\end{array}
$$
while $x,y,\varphi, \eta_1, \eta_2,\eta_3, p_x, p_y,p_\varphi, \dot\eta_3$ are arbitrary.
\medskip

The mechanical Hamiltonian for a skater on the tilted ice-rink with gravitational force reads
$$H_1(x,y,\varphi, \eta_1,\eta_2,\eta_3)=\frac{1}{2m}(\eta_1^2+\frac{1}{k^2}\eta_2^2+\eta_3^2)+\lambda x.$$
The dynamics is composed of all elements of $\sT\sT^\ast M$ that are in relation $\mathcal{D}_C$ with $\xd H_1(\sT^\ast M)$,
which means that
$$z^1=\frac{1}{m}\eta_1,\; z^2=\frac{1}{mk^2}\eta_2,\; z^3=\frac{1}{m}\eta_3=0,\; p_x=\lambda, \; p_y=0,\; p_\varphi=0.$$
It follows then that $\eta_3=0$ and the effective phase space is $C^\ast\subset\sT^\ast M$. The integrability condition is $\dot\eta_3=0$. The integrable part of the dynamics is given by 
$$\begin{array}{lcl}\vspace{10pt}
\displaystyle\dot x= \frac{1}{m}\eta_1\cos\varphi,   &\quad &\dot\eta_1= -\lambda\cos\varphi, \\\vspace{10pt}
\displaystyle\dot y= \frac{1}{m}\eta_1\sin\varphi,   &\quad & \dot\eta_2= 0, \\\vspace{10pt}
\displaystyle\dot\varphi= \frac{1}{k^2 m}\eta_2,
\end{array}$$
which are the same as equations (\ref{eq:33}). 
\medskip

Now, suppose that a point-like charge $q$ is located at a distance $d$ from the center of the skate, as indicated in the figures below, and that a constant magnetic field is applied in the direction perpendicular to the rink. For the choice of the magnetic potential $A = Bx\mathrm{d} y$, with $B$ constant, the Hamiltonian written in terms of coordinates adapted to the constraints is
$$H_2(x,y,\varphi, \eta_1,\eta_2,\eta_3)=\frac{1}{2m}\left(
(\eta_1-Bqx\sin\varphi)^2+\frac{1}{k^2}(\eta_2-Bqdx\cos\varphi)^2+(\eta_3-Bqx\cos\varphi)^2\right).$$
\begin{figure}[h]
    \centering
    \begin{minipage}[t]{0.48\textwidth}
        \begin{center}
\begin{tikzpicture}
\filldraw[gray!20!white] (-2,-1.5) -- (2,-1.5) -- (1.5,-.5) -- (-1.5,-.5) -- (-2,-1.5);
\draw (-2,-1.5) -- (2,-1.5) -- (1.5,-.5) -- (-1.5,-.5) -- (-2,-1.5);
\filldraw[lightgray] (-1,0) arc [radius=1, start angle=180, end angle=360];
\draw[black] (-1,0) arc [radius=1, start angle=180, end angle=360];
\filldraw[blue] (-.5,-.5) circle[radius=0.03];
\draw (-.5,-.5) -- (0,-.5);
\node at (-.25,-.3) {$d$};
\node[left,blue] at (-.5,-.5) {$q$} ;
\draw (-1,0) -- (1,0);
\draw[thick] (0,-1) -- (0,1);
\draw[gray, ->] (-.15,.7) arc [radius=.2, start angle=135, end angle=405];
\filldraw (0,-1) circle [radius=0.03];
\node[right] at (0,-1.2) {$(x,y)$};
\node[right] at (.2,.7) {$\varphi$};
\draw[->] (-1.8,-.7) -- (-1.8,.5);
\node[right] at (-1.8,0) {$B$};
\end{tikzpicture}
\end{center}
    \end{minipage}
    \hspace{0.02\textwidth}
    \begin{minipage}[t]{0.48\textwidth}
       \begin{center}
\begin{tikzpicture}
    \draw[->] (0,0) -- (2.6,0) node[right] {$x$};
    \draw[->] (0,0) -- (0,2.6) node[above] {$y$};
    \draw[thick, gray] (0.5,1) -- ({0.5+2*1.2*cos(40)}, {1+2*1.2*sin(40)});
    \node at (0.5+0.9,1+1.4) {$d$};
      \draw[<->,black] 
        ({\dx + \shiftx}, {\dy + \shifty}) 
        -- 
        ({0.5+2*cos(40) + \shiftx}, {1+2*sin(40) + \shifty});
    \fill[blue] ({0.5+2*cos(40)}, {1+2*sin(40)}) circle (2pt) node[right=2pt, blue] {$q$};
    \fill ({0.5+1.2*cos(40)}, {1+1.2*sin(40)}) circle (2pt);
    \draw (0.5+0.6,1+0) arc[start angle=0, end angle=40, radius=0.6];
    \node at (0.5+0.8,1+0.25) {$\varphi$};
    \draw[dotted] (0.5,1) -- ({0.5+2*1.2*cos(40)}, {1});
\end{tikzpicture}
\end{center}
    \end{minipage}
\end{figure}

\noindent If the charge is positioned at the center, that is when $d=0$, we do not see the effect of a magnetic field because the Lorenz force is perpendicular to the skate and is compensated by the constraint. If the charge is a little off the center, then the Lorenz force influences the rotation of the skate. The condition $z^3=0$ translates to $\eta_3=Bqx\cos\varphi$, and gives the integrability conditions on $\dot\eta_3$. The effective phase space is the affine subbundle of $\sT^\ast M$ defined by $\eta_3=Bqx\cos\varphi$. The rest of the equations read
$$\begin{array}{l}\vspace{10pt}
\displaystyle\dot x= \frac{1}{m}(\eta_1-Bqx\sin\varphi)\cos\varphi,  \\ \vspace{10pt}
\displaystyle\dot y= \frac{1}{m}(\eta_1-Bqx\sin\varphi)\sin\varphi,   \\ \vspace{10pt}
\displaystyle\dot\varphi= \frac{1}{k^2 m}(\eta_2-Bdx\cos\varphi), \\ \vspace{10pt}
\displaystyle\dot\eta_1=\frac{Bq}{m}\cos\varphi\left[\frac{1}{k^2}(x+d\cos\varphi)(\eta_2-Bdqx\cos\varphi)+(\eta_1-Bqx\sin\varphi)\sin\varphi\right] \\\vspace{10pt}
\displaystyle\dot\eta_2= \frac{Bqdx}{mk^2}\sin\varphi\left(\eta_2-Bqdx\cos\varphi\right).
\end{array}$$
Numerical integration produces the following trajectory, for a convenient choice of parameters. The picture shows the projection on the $(x,y)$ plane.
\begin{center}
\includegraphics[scale=0.3]{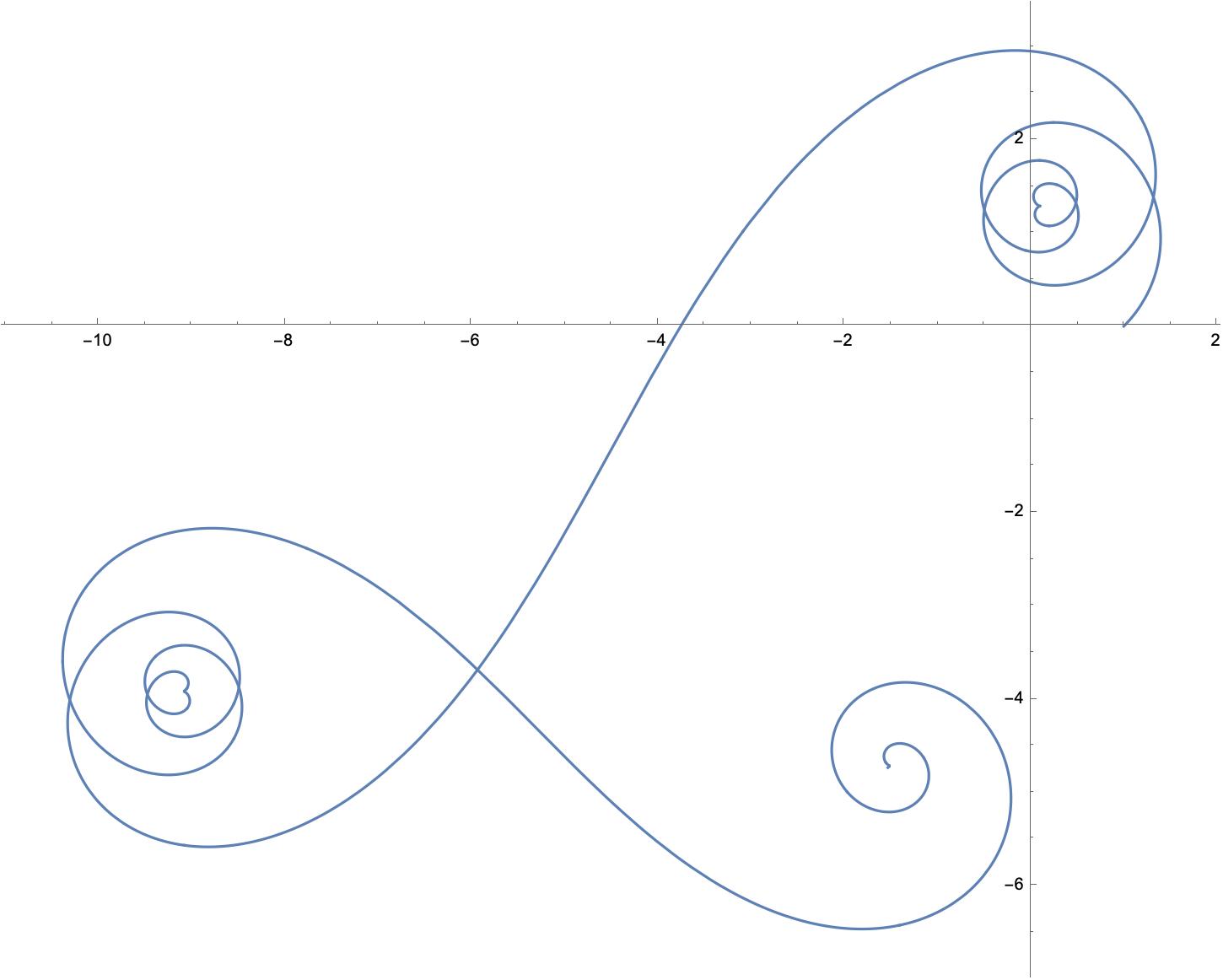}
\end{center}
\end{example}

\begin{example}[A ball on a table] The procedure described in Example \ref{ex:7} can be repeated for any skew algebroid $E$ and any linear constraints $C\subset E$. The resulting Dirac algebroid $D_C$ is a double vector subbundle of the double vector bundle $\sT^\ast E^\ast\oplus_{E^\ast}\sT E^\ast$.

We are going to construct a Dirac algebroid, that can be used to describe the behaviour of a ball rolling on a table without slipping. We can then use any mechanical potential depending on positions, as well as a magnetic-like potential depending linearly on velocities. The initial algebroid structure, before applying the constraints, is a product structure of the canonical Lie algebroid $\sT\R^2\rightarrow \R^2$, which encodes the position and velocity of the center of the ball, and Lie algebra $\mathfrak{so}(3)$, which describes the rotational degrees of freedom. The Lie algebroid $E=\sT\R^2\times \mathfrak{so}(3)$ is a result of a reduction of the full Lagrangian space of infinitesimal configurations, which is $\sT(\R^2\times SO(3))$, with respect to rotational symmetries. This can be done because we consider the case when the Lagrangian and Hamiltonian do not depend on the actual orientation of the ball in the space. The rotational velocities and momenta are, however, important. The Lie bracket on $E$ is given by the Lie bracket of vector fields on $\R^2$ and the usual Lie bracket in the Lie algebra $\mathfrak{so}(3)$. Vector fields on $\R^2$ and constant sections of $\R^2\times\mathfrak{so}(3)\rightarrow \R^2$ commute. The anchor map $\rho$ for this algebroid consists of projection on the first factor.

In the natural basis of sections $(\partial_x,\partial_y,\ell_x,\ell_y,\ell_z)$, where $\ell_i$, with $i\in\{x,y,z\}$, are angular velocities along appropriate axes in $\R^3$, the constraint distribution is spanned by the sections
\begin{equation}\label{eq:24}
f_1=R\partial_{x}+\ell_y, \quad f_2=-R\partial_{y}+\ell_x, \quad f_3=\ell_z.
\end{equation}
We introduce the bundle metric $g$ in $E$, which in the basis dual to $(\partial_x,\partial_y,\ell_x,\ell_y,\ell_z)$ is given by the matrix $\mathrm{diag}(1,1,k^{2},k^{2},k^{2})$, where $k^{2}=\frac{I}{m}$, $I$ is the moment of inertia of the ball, and $m$ is its mass. We assume that the mass density in our ball is spherically symmetric. Using the metric $g$, we choose two additional basic sections,  $f_4$ and $f_5$, perpendicular to each other and to the sections that define the constraint distribution:
\begin{equation}\label{eq:25}
f_4=k^2\partial_x-R\ell_y, \qquad f_5=k^2\partial_y+R\ell_x.
\end{equation}
Denoting by $(\phi^A)$, $A\in\{1,\ldots, 5\}$, the dual basis to $(f_A)$, we can write the bundle metric as
$$g=(R^2+k^2)(\phi^1\otimes\phi^1+\phi^2\otimes\phi^2)+ k^2\phi^3\otimes\phi^3+k^2(k^2+R^2)(\phi^4\otimes\phi^4+\phi^5\otimes\phi^5).$$
The coordinates associated with $(f_A)$ are denoted by $(z^A)$, and split into groups $(z^a)$ along $C$ and $(z^\alpha)$ in perpendicular directions with $a\in\{1,2,3\}$ and $\alpha\in\{4,5\}$. The dual coordinates associated  with $(\varphi^A)$ are denoted by $(\eta_A)$: again, they split into the groups $(\eta_a)$ and $(\eta_\alpha)$. Note that, using the metric $g$ to construct bases and coordinates is not mandatory here - it is purely a matter of convenience.

We encode the Lie algebroid structure of $E\rightarrow \R^2$ in the Dirac structure $D_\Lambda$, i.e., the graph of a map $\Lambda^\#$ associated with the linear Poisson structure on $E^\ast=\sT^\ast \R^2\otimes \mathfrak{so}(3)^\ast$. The Poisson bivector $\Lambda$ is just the sum of the canonical Poisson structure on $\sT^\ast \R^2$ and the canonical  Poisson structure on $\mathfrak{so}(3)^\ast$.

Using the coordinates $(x,y, \eta_A)$ in $E^\ast$, we can define induced coordinates $(x,y,\eta_A;\, a_x, a_y, z^B)$ on $\sT^\ast E^\ast$ and $(x,y,\eta_A;\, \dot x, \dot y, \dot\eta_B)$ on $\sT E$, as well as the composed coordinate system
$$(x,y,\eta_A;\, a_x, a_y, z^B; \dot x, \dot y, \dot\eta_D)$$
in on $\sT^\ast E^\ast\oplus_{\sT\R^2}\sT E^\ast$. The Dirac structure $D_\Lambda$ is given by
$$D_\Lambda=\{(x,y,\eta_A;\, a_x, a_y, z^B;\, \dot x=Rz^1, \dot y=-Rz^2, \dot\eta_B=c^A_{BD}\eta_Az^D-Ra_x+Ra_y\},$$
where $c^A_{BD}$ are the structure functions of the Lie algebroid $E\rightarrow \R^2$ in the basis $(f_A)$ adapted to constraints, i.e., $[f_D,f_B]=c^A_{BD}f_A$. These functions can be easily calculated using formulae (\ref{eq:24}) and (\ref{eq:25}).

The construction of $D_C$ follows the idea from \cite{GRABOWSKA2011}, and is identical to that described in Example \ref{ex:7}. Denoting with $\Delta$ the graph of $\rho$ restricted to constraints
$$E\oplus_{\R^2}\sT\R^2\supset\Delta=\{(u,v): v\in C, u=\rho(v)\},$$
and with $\Delta^\circ$ its annihilator in $E^\ast\oplus_{\R^2}\sT^\ast\R^2$, we define $D_C$ as
$$D_C=\tau_r^{-1}(\Delta)+\Delta^\circ.$$
The map $\tau_r$ is the right projection in a double vector bundle $D_\Lambda$ as in equation (\ref{eq:23}). The addition of elements of $\Delta^\circ$ can be done in the left or the right vector bundle structure on $\sT^\ast E^\ast\oplus_{\sT\R^2}\sT E^\ast$, since $\Delta^\circ$ is a subbundle of the core of the latter double vector bundle. In coordinates, it amounts to putting $z^4=z^5=0$, and setting $\dot\eta_\beta$ free. We get then
$$D_C=\{(x,y,\eta_A;\, a_x, a_y, z^a, z^\alpha=0;\, \dot x=Rz^1, \dot y=-Rz^2, \dot\eta_b=c^D_{bd}\eta_Dz^d-Ra_x+Ra_y, \dot\eta_\beta)\},$$
with free parameters being $x,y,\eta_A;\, a_x, a_y, z^a, \dot\eta_\beta$.

Now, we are ready to play with our Dirac algebroid. For the free Hamiltonian
$$H_0(x,y,\eta_A)=\frac{1}{2m(k^2+R^2)}\left[\eta_1^2+\eta_2^2\right]+\frac{1}{2mk^2}\eta_3^2+\frac{1}{2mk^2(k^2+R^2)}\left[\eta_4^2+\eta_5^2\right],$$
we get the phase equations
$$\dot x=\frac{R}{m(k^2+R^2)}\eta_1,\quad \dot y=\frac{-R}{m(k^2+R^2)}\eta_2,\quad  \dot\eta_1=\dot\eta_2=\dot\eta_3=0,$$
together with the integrability conditions $\eta_4=\eta_5=0$. As expected (see \cite{Grabowska2008}), the trajectory of the ball on the table will be a straight line, the parameters of which are given by the initial conditions.

Let us now assume that there is a point-like charge $e$ located in the middle of the ball, and that the whole system is placed in a constant magnetic field perpendicular to the table. We can use the magnetic potential $A=Bx\mathrm{d}y$ with $B$ being constant. In the basis $(\varphi^A)$ it reads
$$A=-BxR\varphi^2+Bxk^2\varphi^5.$$
The magnetic Hamiltonian is given by the formula
$$H_1(x,y,\eta_A)=\frac{1}{2m(k^2+R^2)}\left[\eta_1^2+(\eta_2+eBxR)^2\right]+\frac{1}{2mk^2}\eta_3^2+\frac{1}{2mk^2(k^2+R^2)}\left[\eta_1^4+(\eta_5-eBxk^2)^2\right].$$
The phase equations are then
$$\begin{array}{l}\vspace{5pt}
\displaystyle \dot x=\frac{R}{m(k^2+R^2)}\eta_1,\\ \vspace{5pt}
\displaystyle \dot y=\frac{-R}{m(k^2+R^2)}\left(\eta_2+eBRx\right),\\ \vspace{5pt}
\displaystyle \dot\eta_1=\frac{-eBR^2}{m(k^2+R^2)}\left(\eta_2+eBRx\right),\\ \vspace{5pt}
\displaystyle \dot\eta_2=0, \\ \vspace{5pt}
\displaystyle \dot\eta_3=0,
\end{array}$$
with the integrability conditions $\eta_4=0$, $\eta_5=Bxk^2$.
One can check that the ball moves along a circle on the table with the radius depending on the initial conditions. We may also add a mechanical potential of choice to our magnetic Hamiltonian $H_1$. We can obtain aesthetically pleasing trajectories  using the harmonic potential:
$$H_2(x,y,\eta_A)=H_1(x,y,\eta_A)+\frac{1}{2}m\Omega^2(x^2+y^2).$$
The equations read then
$$\begin{array}{l}\vspace{5pt}
\displaystyle \dot x=\frac{R}{m(k^2+R^2)}\eta_1,\\ \vspace{5pt}
\displaystyle \dot y=\frac{-R}{m(k^2+R^2)}\left(\eta_2+eBRx\right),\\ \vspace{5pt}
\displaystyle \dot\eta_1=\frac{-eBR^2}{m(k^2+R^2)}\left(\eta_2+eBRx\right)-Rm\Omega^2x,\\ \vspace{5pt}
\displaystyle \dot\eta_2=Rm\Omega^2 y, \\ \vspace{5pt}
\displaystyle \dot\eta_3=0,
\end{array},$$
and, for a suitable choice of parameters, we can get, e.g., the trajectory:
\begin{center}
\includegraphics[scale=0.3]{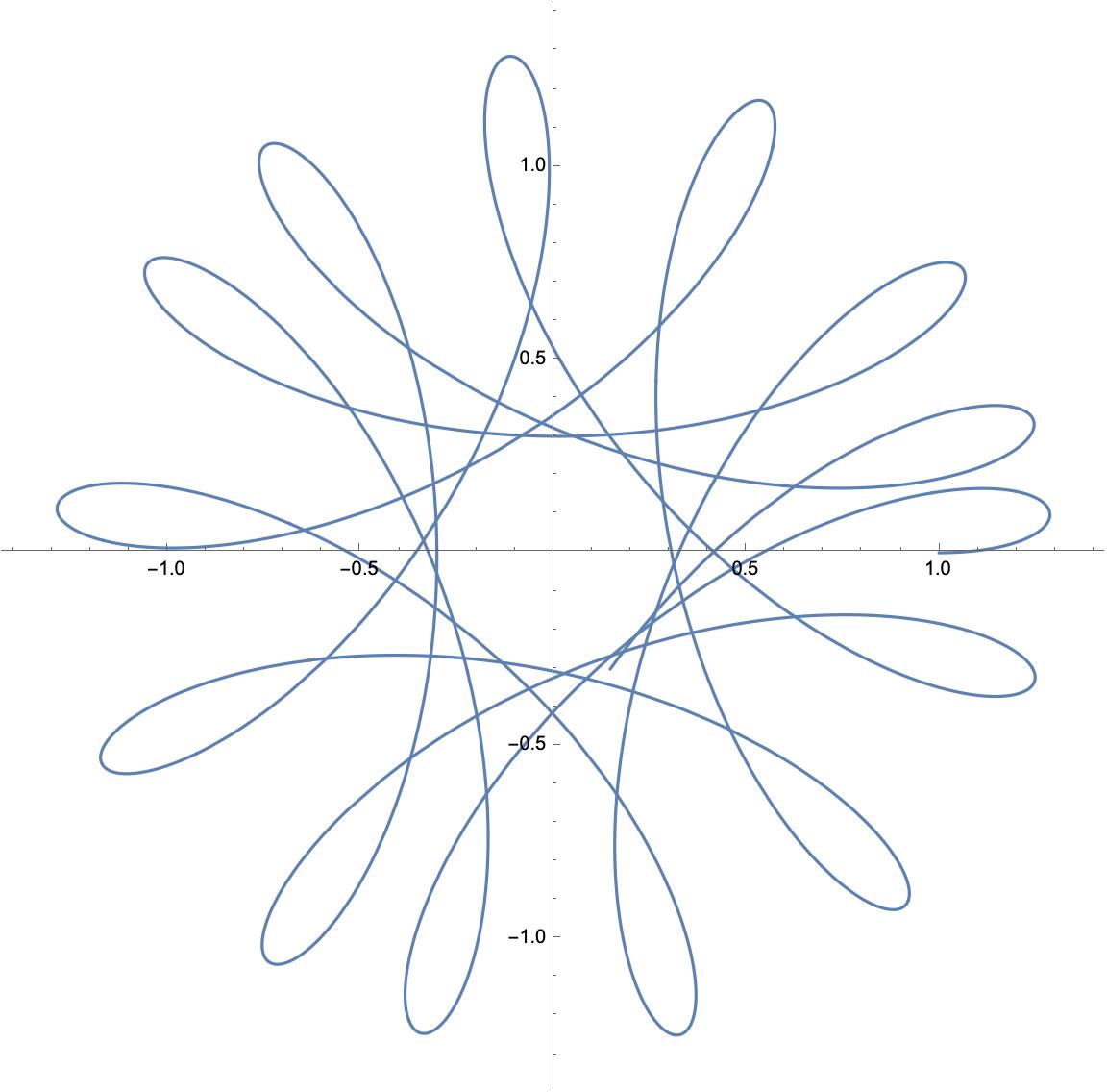}
\end{center}

\end{example}

\section{Summary and perspectives}
In the present note, we have argued that a Dirac algebroid, i.e., a linear almost Dirac structure on a vector bundle, is an effective tool for describing systems with nonholonomic constraints. The structure is defined by the constraints and the underlying geometry of the vector bundle under consideration, and serves for a wide class of Lagrangians. The advantage with respect to skew algebroids or almost Poisson brackets is that the structure is defined by the constraints only and does not require any special form of a Lagrangian or Hamiltonian. The construction, however, is based on the assumption that the constraints are linear. There are other interesting examples in which constraints form an affine subbundle of the Lagrangian configuration bundle. The construction that will, for instance, allow for a phase description of a charged ball in a magnetic field, rolling without slipping on a rotating table is postponed to a separate publication.

\bibliographystyle{acm}
\bibliography{references}

\end{document}